\documentclass[11pt]{article}

\usepackage[margin=1in]{geometry}

\usepackage[utf8]{inputenc}
\usepackage[T1]{fontenc}
\usepackage{amsmath,amssymb}
\usepackage{graphicx}
\usepackage{booktabs}
\usepackage{float}
\usepackage{placeins}
\usepackage{xcolor}

\usepackage{tcolorbox}
\usepackage{authblk}
\usepackage[hidelinks]{hyperref}
\usepackage{cleveref}
\usepackage{enumitem}
\usepackage[nottoc]{tocbibind}

\graphicspath{{Images/}}

\newcounter{prompttemplate}
\crefname{prompttemplate}{Template}{Templates}
\Crefname{prompttemplate}{Template}{Templates}

\newenvironment{promptbox}[1][]{
  \refstepcounter{prompttemplate}%
  \begin{tcolorbox}[
    colback=gray!10,
    colframe=black!50,
    fonttitle=\bfseries,
    coltitle=black,
    title=Template~\theprompttemplate\if\relax\detokenize{#1}\relax\else: #1\fi
  ]}
  {\end{tcolorbox}}

\title{\textbf{A Framework for Using and Evaluating LLMs as Surrogate Experts in Security Surveys: Reliability, Bias, and Implications}}

\author[1]{Despoina Giarimpampa\thanks{Corresponding author: \href{mailto:despoina.giarimpampa@uni.lu}{despoina.giarimpampa@uni.lu}}}
\author[2]{Roland Meier}
\author[1]{Tegawend\'e F. Bissyand\'e}
\author[1]{Vincent Lenders}
\author[1]{Jacques Klein}

\affil[1]{SnT, University of Luxembourg, 6 rue Richard Coudenhove-Kalergi, 1359 Luxembourg, Luxembourg}
\affil[2]{Cyber-Defence Campus, armasuisse Science and Technology (S+T), Feuerwerkerstrasse 39, 3602 Thun, Bern, Switzerland}

\date{}

\begin{document}

\maketitle

\begin{abstract}
\noindent
Expert surveys are widely used in security research to study practitioner workflows and decision-making, yet recruiting domain experts---especially in Security Operations Centres (SOCs), where analysts face high workload, burnout and confidentiality constraints---is difficult and often results in small samples. Large language models (LLMs) offer an appealing alternative by generating synthetic responses at scale, but little guidance exists on when such surrogate participants are reliable. We present a methodological framework for evaluating LLMs as substitutes or supplements to expert survey respondents. Using responses from SOC professionals, we compare persona-based and aggregate LLM-generated answers across multiple models and prompting settings. We measure stability, inter-model agreement and alignment with human responses. Our results show that although LLMs produce internally consistent answers, they systematically diverge from experts, exhibiting reduced variance, central tendency bias and homogenised opinions. This work contributes methodological evidence and practical guidance to the security research community on the appropriate use and limitations of LLM-generated survey responses. We conclude that LLMs are useful for piloting and hypothesis generation but not for replacing expert elicitation, and we discuss implications for researchers using LLM-augmented surveys.

\vspace{0.5em}
\end{abstract}

\section{Introduction}
\label{intro}

Security research relies heavily on expert judgement. Studies of incident response workflows, Security Operations Centre (SOC) analyst decision-making, tooling adoption and organisational maturity all show how much of our empirical understanding of cybersecurity depends on surveys and interviews with human practitioners \cite{axon2020data, mohd2022incident, agyepong2023systematic, abd2021model, kumar2022role, eriksson2022towards, mink2023everybody, akinrolabu2018challenge, ehsan2021expanding, oesch2020assessment, dupont2023tensions, dietrich2018investigating}.

Yet, these surveys face challenges: response rates are low, expert samples are small and the resulting insights are often fragile. Practitioners are difficult to recruit due to time pressure, resource constraints or confidentiality constraints, and many research efforts struggle to obtain sufficient participation in cybersecurity. In some cases \cite{ghazi2018survey}, valuable surveys are delayed or abandoned altogether because too few experts respond. More broadly, meta-analyses find that online surveys yield a weighted mean response rate of only $\sim$44\%, trailing other survey modes by 11–12 percentage points \cite{wu2022response}. These persistent challenges raise the question of how the community can continue to study expert practices when experts themselves are unavailable. Related work points to deeper obstacles: reproducibility and non-response bias in software engineering surveys \cite{ghazi2018survey}, difficulties in capturing nuanced human factors in cybersecurity contexts \cite{ramlo2021human} and design and validity challenges in survey instruments \cite{jansen2023employing}. These problems are particularly acute in SOC environments, where experts face high workload, burnout and staffing shortages that further limit availability for research participation \cite{haney2024towards, thimmaraju2025human}. Thus, there is a clear need for approaches that can support or augment expert elicitation \cite{stanfordSocialScience}.

Meanwhile, because Large Language Models (LLMs) are trained on vast amounts of human-written text, they encode and reflect broad patterns of human knowledge. This capacity has led researchers in other domains to explore whether LLMs can serve as simulated survey respondents, helping to stress-test questionnaires or expand limited datasets \cite{kim2023ai, mburu2025methodological, nie2025data, luo2025llm4sr}. Psychometric studies suggest that LLMs can evaluate survey items with some rigour \cite{liu2025leveraging}, while broader policy and human–computer interaction research highlights both the promise and the risks of AI-augmented surveys, including concerns about validity, reproducibility and trust \cite{haney2024towards, norcExpertView}. 

If feasible, LLM-generated responses could allow researchers to pilot instruments, explore counterfactual populations or prototype studies without immediate access to practitioners. However, prior work cautions that synthetic responses are fraught with risk: models may hallucinate, exaggerate agreement, smooth over disagreement or generate answers that appear plausible but lack grounding in real practice \cite{zhou2024chatgpt, liu2025leveraging}. Recent evaluations show that LLMs often fail to reproduce human-like response variability, with alignment methods such as Reinforcement Learning from Human Feedback (RLHF) shifting distributions in ways that reduce representativeness~\cite{tjuatja2024llms}. Practitioner guidance similarly notes that while LLMs hold promise for coding open-ended responses and assisting in questionnaire design, they fall short in reliably capturing nuanced group-level opinion patterns and raise concerns about reproducibility between model updates \cite{norcExpertView}.

Despite growing interest in LLM-augmented surveys, researchers currently lack methodological guidance on when synthetic participants are appropriate, how their reliability should be assessed and what risks they introduce. In other words, while LLMs are increasingly being used to augment surveys, we lack principled ways to evaluate their validity as surrogate respondents. Because cybersecurity research depends fundamentally on human expertise and practitioner judgement, understanding whether synthetic respondents accurately represent human perspectives is a critical human-centred security question.

\textbf {This paper.} We address this gap from a methodological perspective. Rather than asking whether a particular model performs well, we ask how researchers should responsibly use—or avoid using—LLMs in expert cybersecurity surveys. We propose a reproducible evaluation framework for assessing LLMs as surrogate participants and study how methodological choices—such as prompt design, sampling and aggregation—shape the credibility of synthetic responses. These elements are often treated as technical details; yet, they are central to reproducibility and to understanding when synthetic responses are trustworthy. This framework serves both as a tool for researchers to responsibly use LLMs in survey simulation and as a methodological instrument to evaluate the validity and limitations of synthetic survey responses.

We investigate the credibility of synthetic responses across three complementary settings:  
(1) individual expert simulation through persona-based prompting;  
(2) replication of aggregate, population-level survey distributions; and  
(3) temporal robustness analyses using multi-year SOC survey data.  
This structure allows us to probe not only whether LLMs can mimic expert opinions, but also how stable, reproducible and representative such simulations are under realistic research conditions.

\paragraph{Contributions and Human Factors Implications}
This paper makes four contributions:

\begin{itemize}
    \item We propose a general methodological framework for using and evaluating LLMs as surrogate participants in expert cybersecurity surveys.
    \item We provide an empirical comparison between real SOC experts and LLM-generated respondents across individual, aggregate and temporal settings, including stability and robustness analyses.
    \item We identify systematic failure modes—such as reduced variance, central tendency bias and homogenised opinions—and discuss their implications for the responsible use of LLMs in survey-based security research.
    \item We identify human factors risks associated with synthetic survey respondents, including loss of expert variability and misrepresentation of practitioner perspectives and discuss their implications for preserving validity in human-centred security research.

\end{itemize}

\FloatBarrier
\section{Related work}
\label{Related_work}
There has been growing interest in leveraging artificial intelligence to support or augment survey design, response synthesis and analysis. In the social sciences and psychology, early investigations have explored how AI and LLMs can assist in generating plausible survey responses or enhancing the robustness of questionnaire-based studies \cite{mburu2025methodological, nie2025data}. Similarly, in software engineering, AI-generated responses have been proposed to simulate expert reasoning and support predictive tasks \cite{steinmacher2024can}, partly motivated by persistent challenges such as low response rates and sampling bias.

Despite this growing interest, existing efforts remain fragmented. Prior work demonstrates that LLMs can generate plausible survey responses, but provides limited evidence about how well these responses align with real expert judgements. As a result, researchers have limited methodological guidance on how to validate, benchmark or responsibly use synthetic responses in survey-based research.

In cybersecurity, this gap is even more pronounced. Although recent surveys compile hundreds of LLM-based technical applications in security \cite{zhang2025llms}, they explicitly note the absence of structured expert evaluation.
Recent preprints take initial steps towards using LLMs as surrogate experts in survey contexts \cite{steinmacher2024can, zhou2024chatgpt}. These studies highlight the potential of AI to fill data gaps when traditional methods such as Delphi studies \cite{dalkey1963experimental} or structured interviews are not viable at scale. However, prior work typically provides limited methodological scrutiny. It has primarily evaluated LLM-generated survey responses using limited prompt settings or general-population questionnaires, with less attention to run-to-run stability, cross-model variability or systematic comparison against domain experts.
Consequently, it remains unclear when synthetic responses reflect expert reasoning and when they introduce bias or artificial consensus.

Related work in social cybersecurity explores how AI methods detect and model adversarial activity or online manipulation at scale \cite{mulahuwaish2025survey}. While this line of research focuses on computational detection and automation, our study instead addresses a complementary question: whether LLMs can serve as surrogate participants for knowledge elicitation in security surveys. In the specific context of SOCs, prior LLM research has focused on alert triage, threat summarisation or workflow automation, but not on replacing or augmenting expert responses in survey-based studies.

In summary, prior work demonstrates both the promise and the risks of LLM-augmented surveys, but provides limited evidence about their reliability in expert domains or guidance on how such systems should be evaluated. Our work addresses this gap by systematically assessing LLMs as surrogate participants in cybersecurity surveys and by developing a reproducible evaluation methodology that emphasises stability, alignment with human experts and considerations for responsible use.
\FloatBarrier
\section{Methodology}
\label{methodology}

This section presents a general methodological evaluation framework that formalises how LLM-based survey simulation should be conducted and assessed to ensure validity in human-centred security research. This framework is general and does not depend on the particular evaluation settings presented later. It specifies how LLM-based survey simulation should be structured in principle, while \Cref{expdesign} instantiates the framework in three different empirical contexts that reflect common realities of survey-based security research.

\begin{figure*}[!t]%
\centering
\includegraphics[width=.75\textwidth]{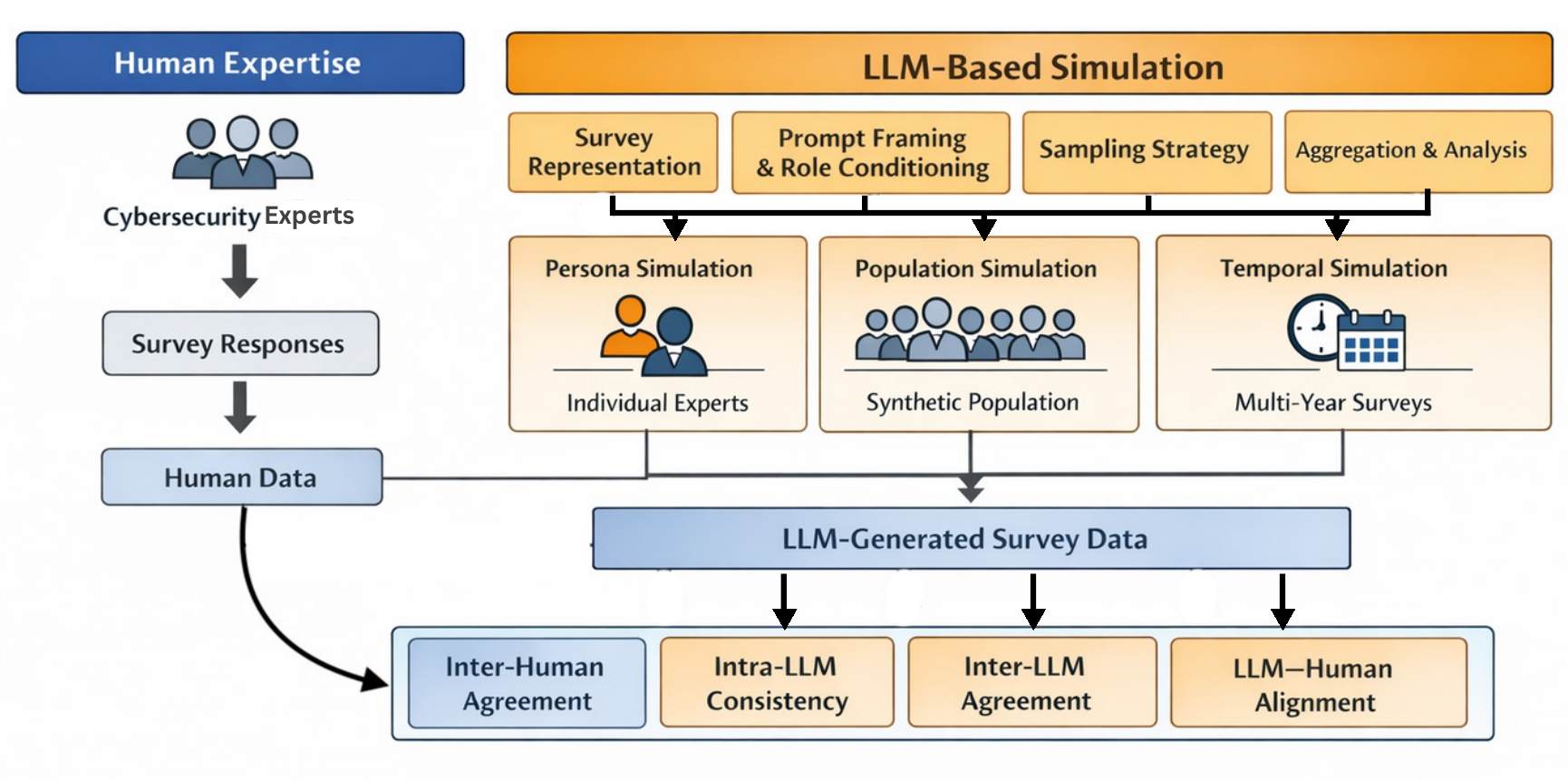}
\caption{Overall framework for evaluating LLMs as surrogate respondents in cybersecurity surveys. Human expert responses and synthetic LLM-generated responses are produced independently and compared across multiple agreement and stability dimensions.}\label{fig:flowchart_soups}
\end{figure*}

\subsection{Conceptual framework for LLM-based survey simulation}

In our methodology, LLMs are queried in three modes: as simulated individual experts, as an aggregated population and as surveys spanning multiple years. The overall framework consists of four components:

\begin{enumerate}
    \item \textbf{Survey representation.} Questions are encoded as categorical, multiple-choice or Likert-scale items with standardised formatting.

    \item \textbf{Prompt framing and role conditioning.} Prompts define the survey task and response format, including persona/distributional data. No example answers are provided.

    \item \textbf{Sampling strategy.} Because LLMs are stochastic, we query each model multiple times under identical conditions. This produces a sample-based representation of the model's behaviour, enabling estimation of stability, dispersion and run-to-run variability.

    \item \textbf{Aggregation and representation.} Repeated outputs are aggregated using survey-appropriate rules to form comparable responses.
\end{enumerate}

These components were selected because they reflect the core methodological steps implicitly performed in prior LLM-based survey simulations. By formalising these steps into a unified framework, we enable systematic evaluation, reproducibility and critical assessment of synthetic survey respondents.

The framework's validity is demonstrated by its ability to reveal systematic differences between synthetic and human responses across multiple survey settings, including individual, aggregate and temporal analyses. By applying consistent methodological components across these contexts, the framework identifies patterns of instability, reduced variability and misalignment that would not be apparent under single-run or single-model evaluations.

\begin{figure}[t]
  \centering
  \includegraphics[width=0.95\columnwidth]{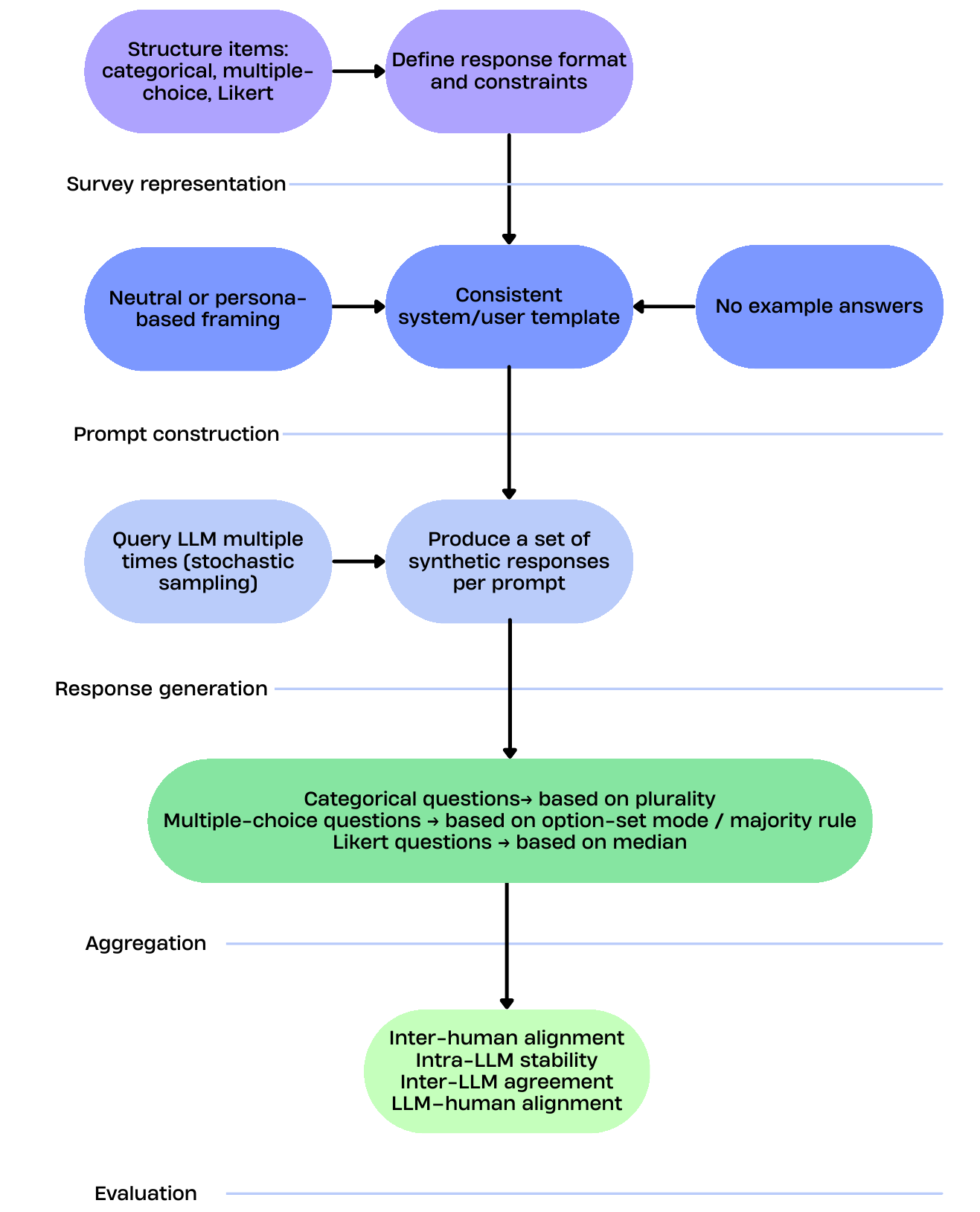}
  \caption{LLM survey simulation framework. Survey questions are represented in structured form, synthetic responses are generated through repeated prompting, aggregated using survey-appropriate rules and evaluated for stability and alignment with human expert responses.}
  \label{fig:flowchart}
\end{figure}

\Cref{fig:flowchart_soups} presents the overall evaluation framework, while \Cref{fig:flowchart} details the LLM simulation pipeline.

\subsection{Prompting principles}

Prompting follows three methodological principles:

\begin{enumerate}

\item \textbf{Task stability:} We use a consistent system--user prompt structure to minimise drift across different model runs.

\item \textbf{Role conditioning:} When simulating personas, prompts contain contextual metadata (e.g. role, industry, organisational size) to promote domain-appropriate reasoning.

\item \textbf{Zero-shot independence:} Prompts do not include human example answers, ensuring that models reveal their own inductive biases.
\end{enumerate}

\subsection{Sampling and stability assessment}

Because LLM outputs are stochastic, using a single generation per model can obscure natural variability in option selection and numerical ratings. To quantify this variability, we propose ten independent samples per model in each setup. This number is motivated by empirical findings in the self-consistency literature \cite{wang2022self}, which show that most stability and accuracy gains occur within the first 5--10 generations, after which improvements plateau.

Repeated sampling enables the analysis of:
(i) intra-LLM stability (repeatability across runs);
(ii) induced response variance; and
(iii) model-level dispersion independent of human data.

\subsection{Aggregation strategies}

We apply question-type-specific aggregation:
\begin{itemize}
    \item \textbf{Categorical:} plurality vote across runs.
    \item \textbf{Multiple-choice:} exact option-set mode, majority rule or Bayesian inclusion rules.
    \item \textbf{Likert-scale:} median across runs to respect ordinal structure.
\end{itemize}

\subsection{Evaluation dimensions}

We evaluate synthetic survey outputs along four dimensions:

\begin{itemize}

\item \textbf{Inter-human agreement} to quantify natural expert variability.

\item \textbf{Intra-LLM consistency} using pairwise similarity across repeated runs for each model, quantifying how stable each model's responses are across various independent generations.

\item \textbf{Inter-LLM agreement} to measure convergence or divergence across models.

\item \textbf{LLM--human alignment} using question-type-specific metrics.

\end{itemize}

These dimensions form the basis for the experimental analyses described in \Cref{expdesign}.

\subsection{Research questions}

To operationalise the framework and evaluate its effectiveness, we address the following research questions:

\begin{itemize}

\item \textbf{RQ1:} To what extent do LLM-generated responses align with individual expert judgements in cybersecurity surveys?

\item \textbf{RQ2:} Can LLMs reproduce aggregate population-level response distributions observed in practitioner studies?

\item \textbf{RQ3:} How stable and sensitive are synthetic responses across runs, models and survey settings?

\item \textbf{RQ4:} To what extent do LLMs generalise to expert surveys conducted outside their training horizon, and do temporal alignment patterns suggest limitations in generalisation?

\item \textbf{RQ5:} What systematic differences or failure modes emerge when using LLMs as surrogate experts?

\item \textbf{RQ6:} What methodological implications emerge for researchers considering the use of LLM-generated responses?

\end{itemize}
\FloatBarrier
\section{Experimental design}
\label{expdesign}

Building upon \Cref{methodology}, we instantiate the framework in three evaluation settings. 

\begin{itemize}
    \item \textbf{Persona-based expert simulation (Setting 1).}
    We collect responses from real SOC professionals and compare them directly with LLMs prompted to simulate matching expert personas.

    \item \textbf{Aggregate distribution replication (Setting 2).}
    We evaluate whether LLMs can reproduce population-level response distributions when only summary statistics from practitioner studies are available.

    \item \textbf{Temporal robustness analysis (Setting 3).}
    We compare LLM-generated responses with multi-year practitioner surveys to assess temporal alignment and examine whether models generalise to survey data beyond their training horizon.
\end{itemize}

Each setting reflects a distinct goal and a different way of using LLMs, but all share the same prompting, sampling, aggregation and evaluation procedures described below.

\paragraph{Environmental setup}
We implemented model querying and data handling using Python scripts on an Ubuntu Linux 24.04 ARM64 virtual machine. We accessed each LLM through its respective API using standard configuration settings (e.g. default temperature). The reason we used each model's default generation parameters was to reflect typical usage conditions and to avoid tuning-induced bias. Sampling temperature is known to substantially affect output diversity and surface similarity in neural language models \cite{caccia2018language}. Tuning this parameter to maximise agreement would implicitly optimise generation towards the evaluation metric rather than assessing intrinsic model behaviour. Finally, we applied prompting templates, including detailed system instructions and contextual framing, uniformly across all models.

\paragraph{Selection of LLMs}
We evaluate our framework by using six LLMs: GPT-4o, GPT-4, DeepSeek, Llama~3.1--8B, Llama~3.2--3B and Gemini Flash. This selection spans multiple model families and includes both proprietary and open-weight architectures. We also include GPT-3.5-turbo in the last setting because its earlier training cut-off in 2021 \cite{Wang2025_llm-knowledge-cutoff-dates} provides a natural contrast for temporal generalisation and potential training-data exposure.

For all models, we disable browsing, retrieval, tool use and external plug-ins. All responses are generated entirely from the model's internal parameters. We called only the standard \texttt{/chat/completions} or equivalent API endpoints.

\subsection{Setting 1: individual expert responses}

\textbf{Scenario:} individual expert elicitation with a small number of practitioners.
\textbf{Goal:} assess whether LLMs can simulate individual security experts and reproduce expert-level variability.
\label{sec:exp1-methods}

\subsubsection{Survey design}
We designed a structured survey to investigate the adoption of AI in SOCs, and administered it to cybersecurity professionals working in SOC environments. The survey instrument is provided in \Cref{app:survey}.
\textbf{Ethical approval for this study was granted by the University of Luxembourg Ethics Review Panel (ERP 23-049 ATSO).}

The questionnaire included categorical (single-choice), Likert-scale (1--5) and multiple-choice (select-all-that-apply) questions. These questions targeted key dimensions of SOC AI integration: the extent of AI deployment, the automation of security workflows, changes in operational performance metrics and perceived challenges or barriers to AI adoption.

To improve clarity and comprehension, we piloted the survey with an internal reviewer, following survey-quality guidelines from the literature \cite{presser2004methods, schaeffer2003science}.

\subsubsection{Human expert responses}
We recruited experts through professional networks and industry contacts. Participants completed the survey independently without priming or example answers. No monetary compensation was provided; participation was voluntary and motivated by professional interest.

\begin{table}[!t]
\caption{Summary of human expert respondents\label{tab:experts}}%
\begin{tabular*}{\columnwidth}{@{\extracolsep\fill}ll@{\extracolsep\fill}}
\toprule
Characteristic & Distribution \\
\midrule
Roles & 2 Managers, 2 Analysts, 2 Specialists \\
Experience & 2 (1--3 yrs), 4 (5--10 yrs) \\
Education & 4 Master's, 1 Bachelor's, 1 PhD \\
Sectors & 2 Healthcare, 1 Finance, 1 Government, 1 Energy, 1 Telecom \\
SOC size & 9--25 members \\
Countries & Multiple countries \\
\bottomrule
\end{tabular*}
\end{table}

We collected responses from six cybersecurity professionals representing diverse roles, expertise levels, industries and organisational contexts. \Cref{tab:experts} summarises in aggregated form their demographic and organisational characteristics. These responses provide a human population against which we compare synthetic outputs.

\begin{promptbox}[System prompt] \label{sysprompt}
Imagine you are a population of x Security Operations Centre experts.

Your assignment is to answer the questions of the following questionnaire regarding AI adoption/false positives in SOCs, reflecting how this population would answer the questions. You are representing diverse viewpoints within this group.

Background information and demographic details are provided for you to assume the roles of various individuals within the population: \texttt{[...]}

Please answer each of the following questions. Use a number for each expert and the prefix for each question (Q\textless number\textgreater) so that it’s clear which answer corresponds to each question.
\end{promptbox}

\subsubsection{Prompt strategy}
We tasked each of the six evaluated LLMs with simulating six expert personas using a consistent system-and-user prompting structure across all models, as shown in \Cref{sysprompt,usprompt,sansprompt}. Our prompting methods include:
(1) \emph{prompt sandwiching} to maintain consistent framing and task stability \cite{subbanarasimha2024sandwich};
(2) \emph{persona-based prompts}, where each model simulates a specific human expert to encourage domain-specific and contextual realism \cite{olea2024evaluating}; and
(3) \emph{demographic-specific cues} based on human participant metadata \cite{haney2024towards}.

Since the prompts provide context but no example answers, and since no chain-of-thought reasoning, fine-tuning or few-shot examples were used, each model's zero-shot behaviour could be evaluated independently. The choice to simulate x experts mirrors the number of real-world respondents in each experiment, allowing direct persona mapping and comparative analysis across both human and synthetic responses in Setting~1 and in the aggregate distributional replication in Settings~2 and~3.

\begin{promptbox}[User prompt] \label{usprompt}
 
\textcolor{blue}{Categorical questions:} (Q\textless number\textgreater) \textless Question\textgreater?  
Options: <given options>, I’d rather not say, Unknown\ I don't know.  
<extra details if needed>

\textcolor{blue}{Multiple-choice questions:} (Q\textless number\textgreater) \textless Question\textgreater?  
Options (choose all that apply): \textless given options\textgreater, I’d rather not say, Unknown\ I don't know.
[extra details if needed]

\textcolor{blue}{Likert questions:}(Q\textless number\textgreater) Rate \textless question\textgreater on a scale from 1 to 5, where 1 is the least and 5 is the most challenging for each item:
[sub-questions\ items]

\end{promptbox}

\begin{promptbox}[SANS prompt] \label{sansprompt}

Imagine you are a population of X Security Operations Centre (SOC) experts.
Your assignment is to answer the questions of the following questionnaire
regarding SOC architectures, capabilities, metrics and performance in SOCs,
reflecting how this population would collectively respond. You represent
diverse viewpoints within this group.

Background information and demographic details are provided for you to assume
the roles of various individuals within the population: \texttt{[...]}

Please answer each of the following questions. Every question has a different
number of responders, with the format:\\
\textbf{(Q\textless number\textgreater) (\textless number of responders\textgreater) \textless question\textgreater.}\\
Use percentages for every option and retain each question prefix (e.g.,
\emph{Q\textless number\textgreater}) so that responses clearly map to each question.
\end{promptbox}

\subsubsection{Repeated runs and data collection}
We executed each model independently in ten separate runs with identical survey inputs, yielding 60 simulated responses per model (6 personas $\times$ 10 runs).

\subsubsection{Aggregation strategy}
We aggregated repeated runs using question-type-specific rules that reflect the structure of the survey items:

\begin{itemize}
    \item \textbf{Categorical questions:}
    we apply plurality voting, selecting the most frequent response across the ten runs.

    \item \textbf{Multiple-choice questions:}
    for each persona and question, we aggregate the ten run-level option sets using three rules:
    (i) an exact option-set mode, which selects the most frequently generated bundle of options across runs;
    (ii) a per-option majority rule, which includes an option if it appears in at least 50\% of runs; and
    (iii) a Bayesian per-option rule with a uniform $\mathrm{Beta}(1,1)$ prior, where options are included if their posterior mean selection probability exceeds a fixed threshold (0.5 in our setups) \cite{ho2016eliciting}.
    We use the exact option-set mode for the main analyses and report majority and Bayesian variants as checks.

    \item \textbf{Likert-scale items:}
    we aggregated at the persona level using the median (which respects the ordinal scale and is robust to outliers) of the ten responses, and reported mean and standard deviation as descriptive summaries.
\end{itemize}

\subsubsection{Analytical metrics}
We evaluated agreement and variability across human and LLM responses using question-type-appropriate metrics. These metrics support both within-group (intra-LLM, inter-human) and cross-group (inter-LLM, LLM--human) comparisons.

\begin{itemize}
    \item \textbf{Categorical: exact match rate (normalised Hamming similarity).}
          For $N$ categorical items with reference responses $y_i$ and corresponding model responses $\hat{y}_i$, the exact match rate is defined as
          \begin{equation}
              \mathrm{EM} = \frac{1}{N} \sum_{i=1}^{N} \mathbb{I}(y_i = \hat{y}_i),
              \label{eq:exact-match}
          \end{equation}
          where $\mathbb{I}(\cdot)$ denotes the indicator function. For single-label categorical variables, this metric is equivalent to normalised Hamming similarity \cite{hamming1950error}.

    \item \textbf{Multiple-choice: Jaccard similarity and Hamming similarity.}
          Let $A_i$ and $B_i$ denote the sets of selected options for item $i$. The Jaccard similarity \cite{jaccard1901etude} is defined as
          \begin{equation}
              J(A_i, B_i) = \frac{|A_i \cap B_i|}{|A_i \cup B_i|}.
              \label{eq:jaccard}
          \end{equation}
          Additionally, selections are encoded as binary vectors $\mathbf{a}_i, \mathbf{b}_i \in \{0,1\}^{|\mathcal{O}|}$, where $\mathcal{O}$ is the option universe. The normalised Hamming similarity over option vectors is defined as
          \begin{equation}
              H(\mathbf{a}_i, \mathbf{b}_i) =
              1 - \frac{1}{|\mathcal{O}|}
              \sum_{j=1}^{|\mathcal{O}|} \left| a_{ij} - b_{ij} \right|.
              \label{eq:hamming-similarity}
          \end{equation}

    \item \textbf{Likert: median absolute differences.}
          Let $x_{ij}$ denote the Likert response of rater $j$ for item $i$, and let $\tilde{x}_{ig}$ denote the median response for persona or group $g$. For two personas or groups $g$ and $g'$, the absolute difference is computed as
          \begin{equation}
              \Delta_i = \left| \tilde{x}_{ig} - \tilde{x}_{ig'} \right|.
              \label{eq:median-absolute-difference}
          \end{equation}
          Differences are summarised descriptively across items, such as mean, median and interquartile range, and overall distributional properties are additionally inspected qualitatively.
\end{itemize}

\subsection{Setting 2: distributional surveys}

\textbf{Scenario:} population-level surveys where only aggregate distributions are available.
\textbf{Goal:} evaluate whether LLMs can reproduce population-level patterns in practitioner surveys.

\subsubsection{Design}
Unlike the previous setup, which simulates individual respondents, the current one elicits population-level judgements directly. We simulate a previous study on SOC alarm triage \cite{alahmadi202299}, which reports aggregate response percentages across 20 practitioners.

\subsubsection{Data source}
The source study reports categorical, multiple-choice and Likert-scale distributions. Where percentages were missing in the published paper \cite{alahmadi202299}, we used complementary details from the author's doctoral thesis \cite{alahmadi2019malware}.

\subsubsection{Prompting and generation strategy}
We created our prompts to instruct each LLM to generate a synthetic population of respondents whose answers, in aggregate, reflect distributions comparable to the original study. Each model simulated a fixed group size matching the reported sample size. We ran each model ten times under identical prompts (following the same \Cref{sysprompt,usprompt}), producing ten synthetic populations per model.

\subsubsection{Analytical metrics}
We compared model-generated distributions with the human reference using two metrics:

\begin{itemize}
    \item \textbf{Jensen--Shannon divergence (JSD):} a symmetric and bounded measure of similarity between two probability distributions $P$ and $Q$ \cite{menendez1997jensen}:
          \begin{equation}
              \mathrm{JSD}(P \parallel Q)
              = \tfrac{1}{2} D_{\mathrm{KL}}\!\left(P \parallel M\right)
              + \tfrac{1}{2} D_{\mathrm{KL}}\!\left(Q \parallel M\right),
              \label{eq:jsd}
          \end{equation}
          where $M = \tfrac{1}{2}(P + Q)$ is the mixture distribution.

    \item \textbf{Chi-square divergence:} also known as Pearson's $\chi^2$ divergence, it quantifies squared relative differences between two distributions $P$ and $Q$ \cite{nielsen2013chi}:
          \begin{equation}
              \chi^2(P \parallel Q) = \sum_{i} \frac{\big(P_i - Q_i\big)^2}{Q_i}.
              \label{eq:chi-square-divergence}
          \end{equation}
\end{itemize}

We applied these measures to all categorical, multiple-choice and Likert-scale questions for which full response distributions were available. For interpretability, median values are used as the primary descriptive statistic for Likert items. (Mode values were conceptually considered but not used in the main analyses.) The repeated runs allow us to estimate the variance of the LLM-generated distributions around the human reference.

\subsection{Setting 3: temporal simulation of surveys}

\textbf{Scenario:} multi-year practitioner surveys on similar topic.
\textbf{Goal:} to assess whether LLM-generated survey distributions track temporal changes in practitioner survey results, or instead remain largely time-insensitive.

\subsubsection{Design and data source}
We extended the distributional simulation to a temporal setting. We use annual SANS SOC survey reports from 2017 to 2025 \cite{sansWhitePapers}, which summarise responses from SOC practitioners on topics such as tooling, staffing, automation and alert handling.\footnote{The 2020 SANS SOC survey report does not exist in the repository; we therefore analyse all available years except 2020.} The surveys report aggregate percentages per response option for a range of categorical, multiple-choice and Likert items.

\subsubsection{Temporal generalisation rationale}
This setting enables us to analyse whether LLM-generated responses align with practitioner survey distributions across time, including surveys conducted before and after model training cut-offs. Specifically, we examine whether models (i) exhibit similar alignment for pre-cut-off survey years, (ii) show reduced alignment for post-cut-off years or (iii) produce time-insensitive response patterns that do not track changes in practitioner survey data.

We include GPT-3.5-turbo in this experiment because its earlier training cut-off in 2021 provides a temporal boundary.

\subsubsection{Prompting and generation strategy}
For each survey year and each question, we instructed the models to generate a synthetic population of SOC practitioners whose aggregate responses reflect the state of SOC practice in that year. We provided a short description of the SANS SOC context and the response scale, and then asked the model to output a fixed number of synthetic responses that can be aggregated into a distribution; see \Cref{sansprompt}.

Each model (the six main LLMs plus GPT-3.5-turbo) generated simulated responses for each year in the 2017--2019 and 2021--2025 surveys. As previously, we ran each model ten times per year, yielding ten synthetic distributions per model-year pair.

\subsubsection{Analytical metrics}
We evaluated temporal alignment using the same divergence measures for distributional surveys, namely JSD and Pearson's $\chi^2$, as defined above. Following the concept drift literature, where temporal change is quantified by tracking distributional statistics over time \cite{gama2014survey, webb2016characterizing}, we consider:

\begin{itemize}
    \item \textbf{Per-year alignment divergence:}
    for each model $m$ and survey year $t$, let $P_{m,t}$ denote the model-induced response distribution and let $H_t$ denote the corresponding SANS reference distribution. Alignment is measured as
    \begin{equation}
        A_{m,t}^{(D)} = D\!\left(P_{m,t} \parallel H_t\right),
        \label{eq:per-year-alignment-divergence}
    \end{equation}
    where $D(\cdot\parallel\cdot)$ is instantiated as either Jensen--Shannon divergence or Pearson's $\chi^2$ divergence.

    \item \textbf{Year-to-year drift in alignment:}
    drift is operationalised as the temporal difference of the alignment divergence across consecutive survey years,
    \begin{equation}
        \Delta A_{m,t}^{(D)} = A_{m,t}^{(D)} - A_{m,t-1}^{(D)}.
        \label{eq:year-to-year-drift}
    \end{equation}
\end{itemize}

By comparing pre- and post-cut-off years for GPT-3.5-turbo and contrasting its behaviour with more recent models (e.g. GPT-4o, Gemini-Flash, DeepSeek), we assess whether observed alignment patterns are consistent with temporal generalisation or suggest spurious memorisation of survey data.
\FloatBarrier
\section{Results}
\label{Results}

We present results for the three setups separately, followed by a cross-experiment synthesis.
For visualisation consistency, all metrics are linearly or monotonically rescaled to a common $[-1,1]$ range, where $+1$ indicates maximal agreement and $-1$ indicates maximal divergence; this normalisation does not alter the underlying metrics.

\subsection{Individual responses and simulation}

In this setup,
we focus on agreement patterns, stability and alignment, but we simplify the analysis to avoid unnecessary question-level detail and instead present aggregated, interpretable trends.

\begin{figure}[H]
  \centering
  \includegraphics[width=1\linewidth]{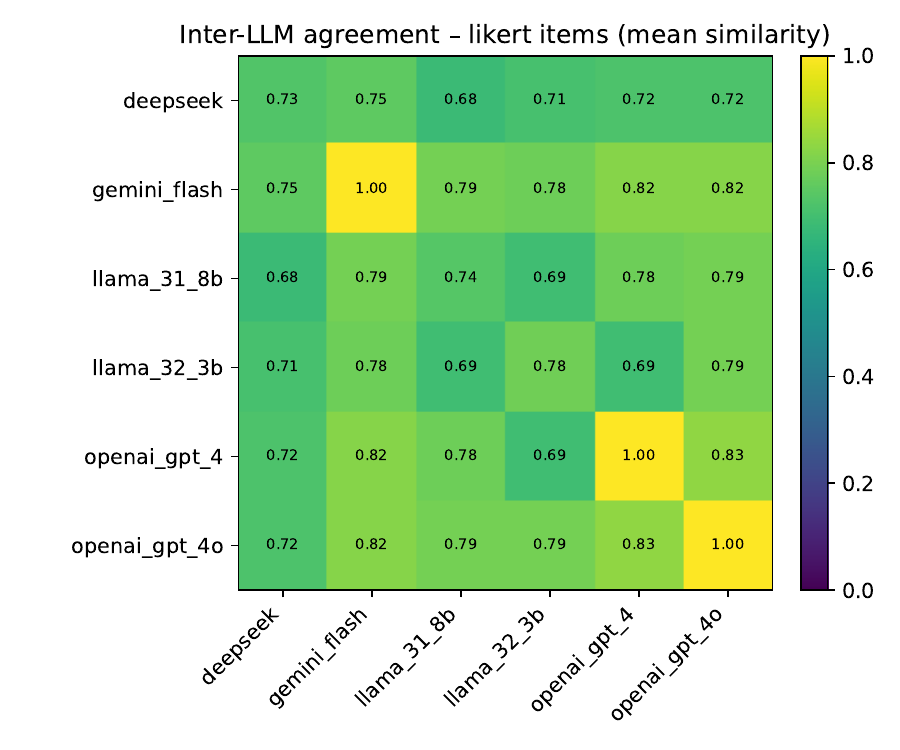}
  \caption{Random inter-LLM similarity for Likert-scale items. Off-diagonal similarities remain well below empirically observed agreement levels, providing a lower-bound benchmark.}
  \label{fig:random_likert}
\end{figure}

\begin{figure}[H]
  \centering
  \includegraphics[width=1\linewidth]{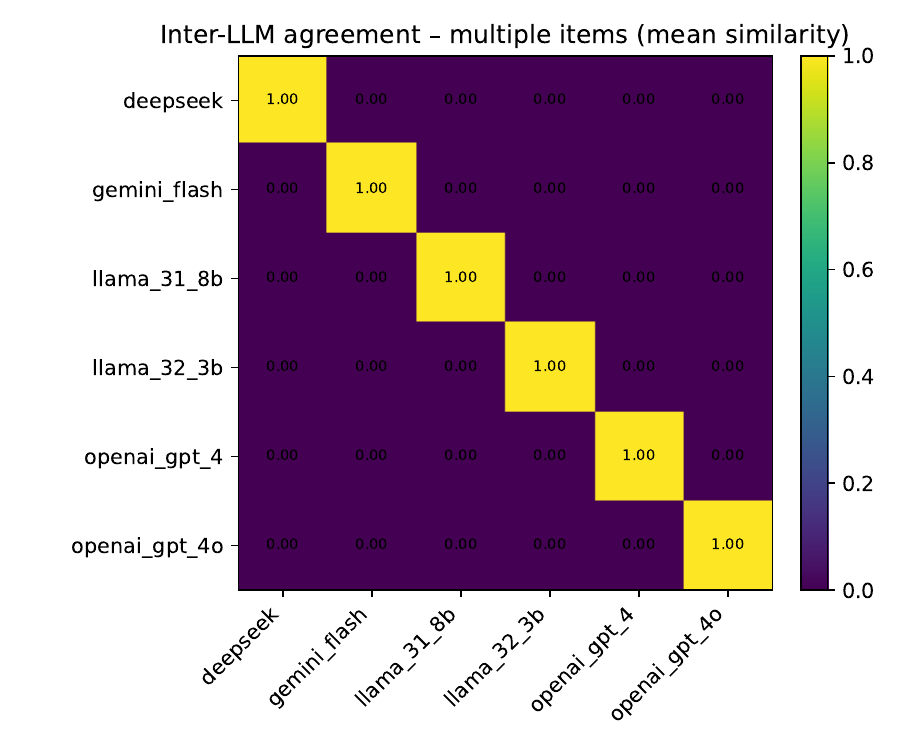}
  \caption{Random inter-LLM similarity for multiple-choice items. Only self-comparisons reach $1.0$, while all cross-model similarities collapse to zero.}
  \label{fig:random_multiple}
\end{figure}

\begin{figure}[t]
  \centering
  \includegraphics[width=1\linewidth]{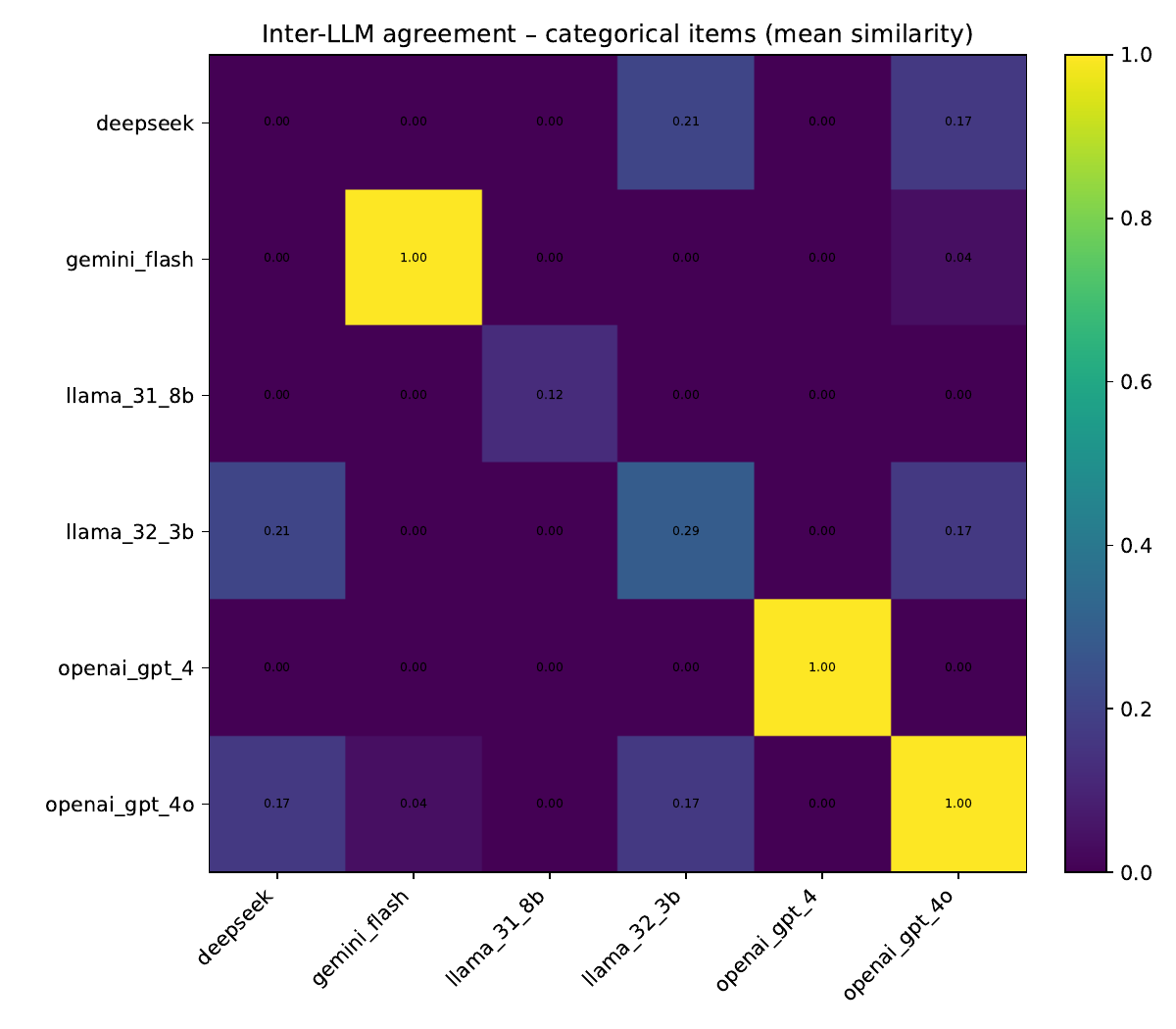}
  \caption{Random inter-LLM similarity for categorical items. The absence of off-diagonal agreement confirms that the metric does not induce chance alignment.}
  \label{fig:random_open}
\end{figure}

\paragraph{Interpreting the similarity metric}
Because our similarity metric ranges from $0$ to $1$, it is essential to interpret values relative to an empirical benchmark rather than treating $0$ as ``good'' and $1$ as ``bad''. To this end, we construct a baseline by computing pairwise similarities between randomly permuted responses of the same question type and scale. This baseline, depicted in \Cref{fig:random_likert,fig:random_multiple,fig:random_open}, approximates the level of agreement that can be expected purely by chance, given the structure of the questionnaire.

\begin{figure}[t]
    \centering
    \includegraphics[width=1\linewidth]{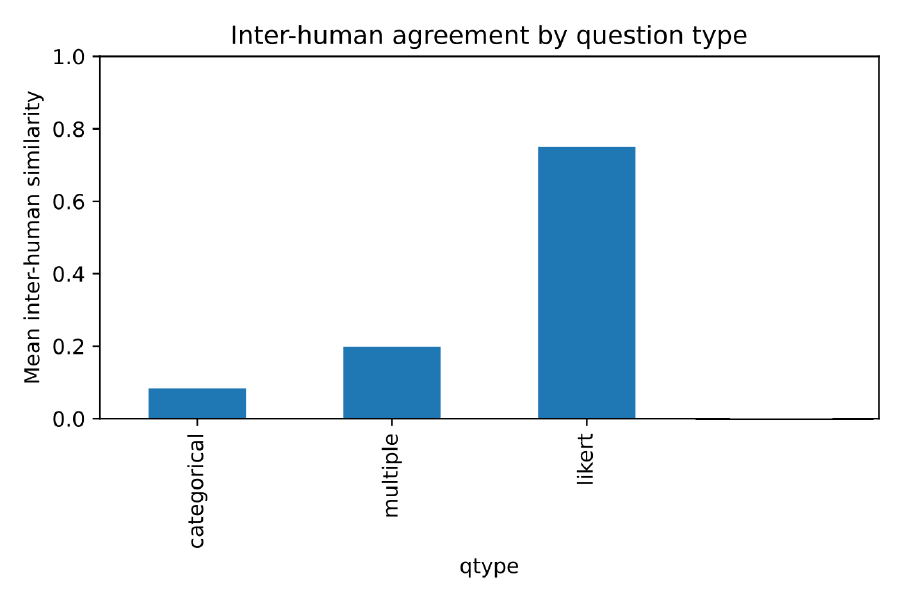}
    \caption{Inter-human agreement by question type.
    Similarity values for categorical, multiple-choice, Likert
    and open-ended items provide an empirical baseline for interpreting
    LLM similarity scores.}
    \label{fig:interhuman-by-qtype}
\end{figure}

\paragraph{Inter-human agreement}
Across the six human experts, responses exhibit substantial dispersion for all question types as shown in \Cref{fig:interhuman-by-qtype}. Categorical items split across multiple plausible options, and multiple-choice items produce heterogeneous option sets. This confirms that human experts do not provide a single consensus.

\begin{figure}[t]
  \centering
  \includegraphics[width=1\linewidth]{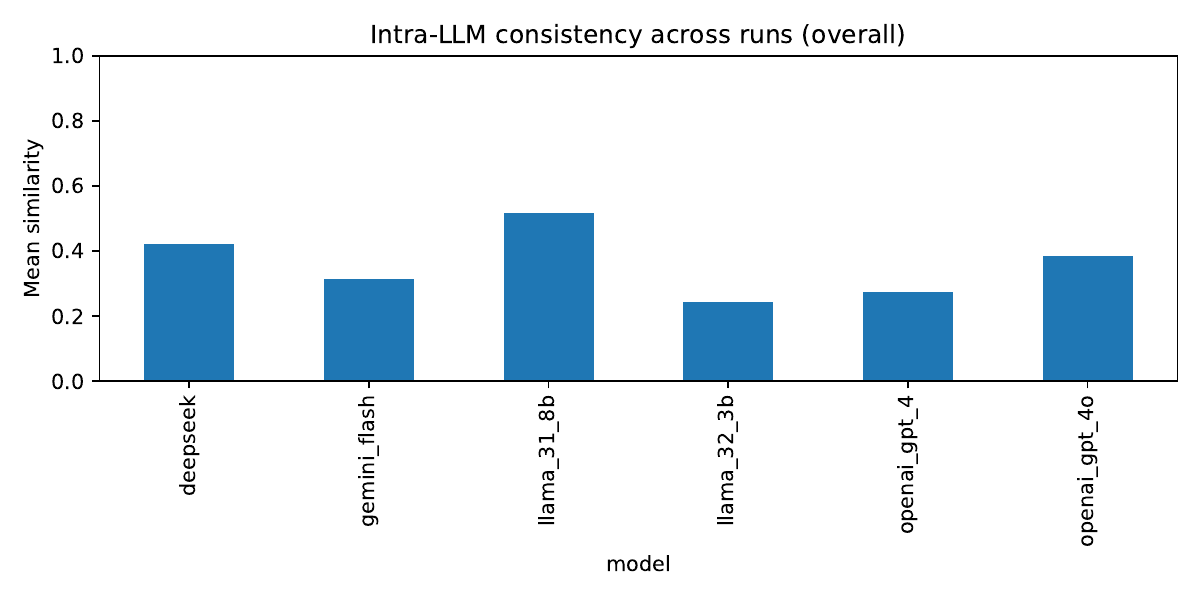}
  \caption{Intra-LLM consistency across runs, aggregated over structured question types. Higher values indicate greater self-consistency under repeated persona-conditioned generations.}
  \label{fig:intra-llm-overall}
\end{figure}

\paragraph{Intra-LLM consistency}
We first assess whether a single model produces stable responses when conditioned on the same expert persona and prompted multiple times. \Cref{fig:intra-llm-overall} reports the overall intra-LLM agreement across runs, aggregated over structured question types (categorical, multiple-choice and Likert). A detailed breakdown by question type is provided in \Cref{fig:intra-llm-merged}. These results indicate that repeated sampling under identical persona-conditioned prompts leads to only partial overlap in generated responses. Even the most stable models fall substantially short of perfect reproducibility, highlighting the persistent impact of stochasticity in LLM generation. Consequently, intra-LLM consistency should be interpreted as a measure of \emph{relative stability} rather than strong determinism: repeated runs reduce variability to some extent, but do not yield near-identical outputs.

\begin{figure}[t]
  \centering
  \includegraphics[width=1\linewidth]{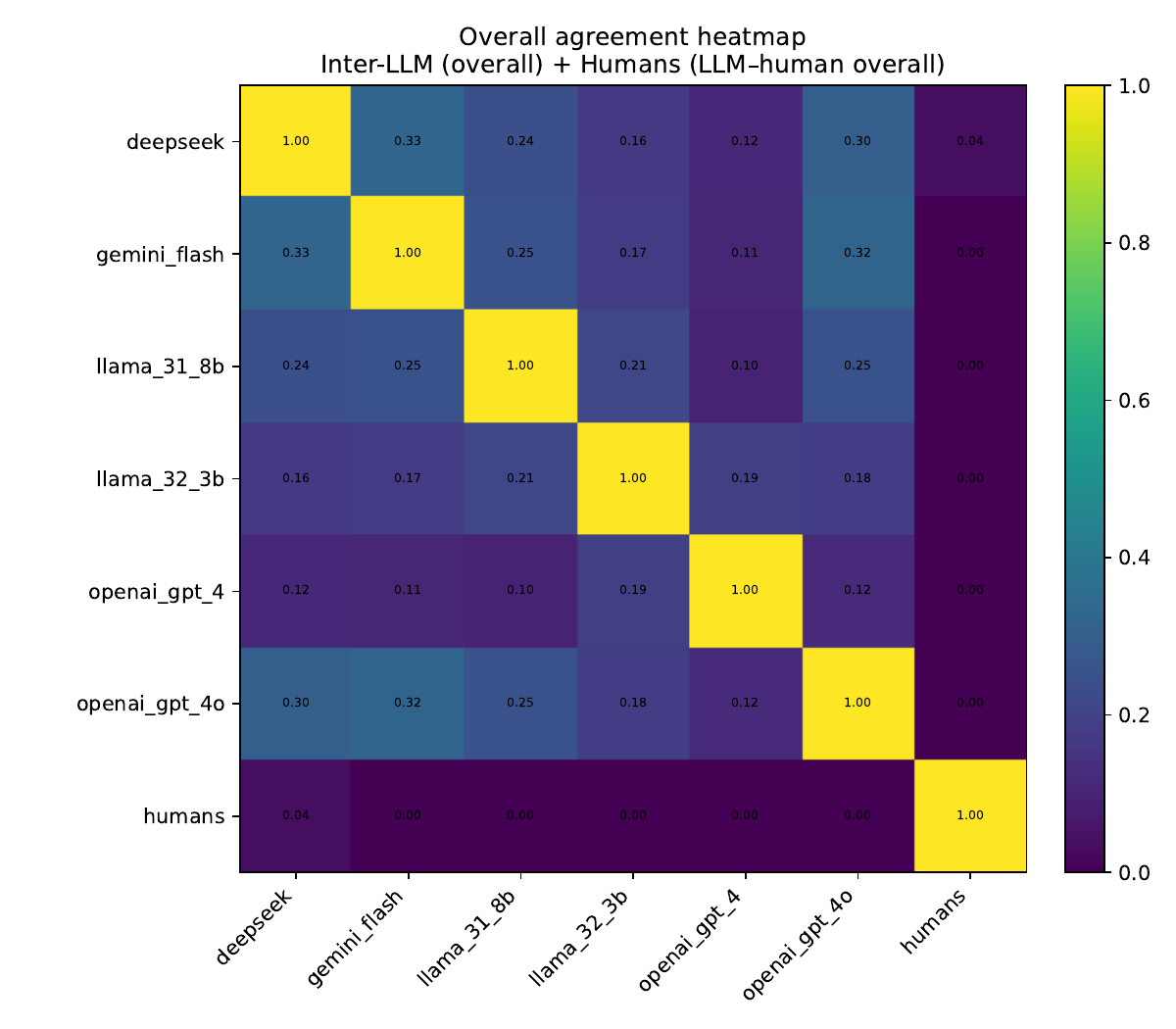}
  \caption{Overall agreement heatmap showing inter-LLM agreement and alignment with human expert responses. The \texttt{humans} row and column report each model's overall similarity to the human experts.}
  \label{fig:overall-interllm-plus-humans}
\end{figure}

\paragraph{Inter-LLM agreement and alignment with humans}
To assess whether different models converge on similar simulated responses, \Cref{fig:overall-interllm-plus-humans} presents the overall inter-LLM agreement heatmap. Each cell represents the mean similarity between two models' simulated responses across all questions, while the additional ``humans'' row and column report each model's overall similarity to the human expert responses.

The heatmap reveals that overall inter-LLM agreement is uniformly low. This indicates that different models rarely converge on similar simulated responses, even when conditioned on the same expert personas and prompts. In addition, similarities between LLMs and human experts are likewise low, and in most cases zero, showing comparable magnitude to inter-LLM similarities.

To summarise, these results show no evidence of meaningful convergence---either across models or towards human expert responses. Agreement among LLMs is weak, and where it exists, it does not exceed alignment with humans. Rather than reflecting shared reasoning or latent expert structure, the observed similarities are consistent with largely independent and noisy response patterns across models.

\paragraph{Overall findings}
Together, \Cref{fig:intra-llm-overall,fig:overall-interllm-plus-humans} reveal a clear separation between reproducibility and alignment. LLMs exhibit limited self-consistency across runs and only moderate agreement with one another, while remaining weakly aligned with human expert responses at an aggregate level.

These findings indicate that persona-conditioned LLM simulation produces responses that are neither fully reproducible nor human-equivalent. Instead, individual simulations should be understood as stochastic samples from model-specific response distributions. As such, LLM-based expert simulation is best suited for exploratory analysis and hypothesis generation, rather than for substituting or closely approximating individual expert judgements.

\subsection{Distributional survey simulation}
\label{sec:exp2}
In the second setup, we evaluated whether LLMs can reproduce aggregate-level survey distributions. This task required each model to generate a population-level response distribution. The comparison to human data was conducted using JSD and Chi-square distance.

\begin{figure}[t]
    \centering
    \includegraphics[width=1\linewidth]{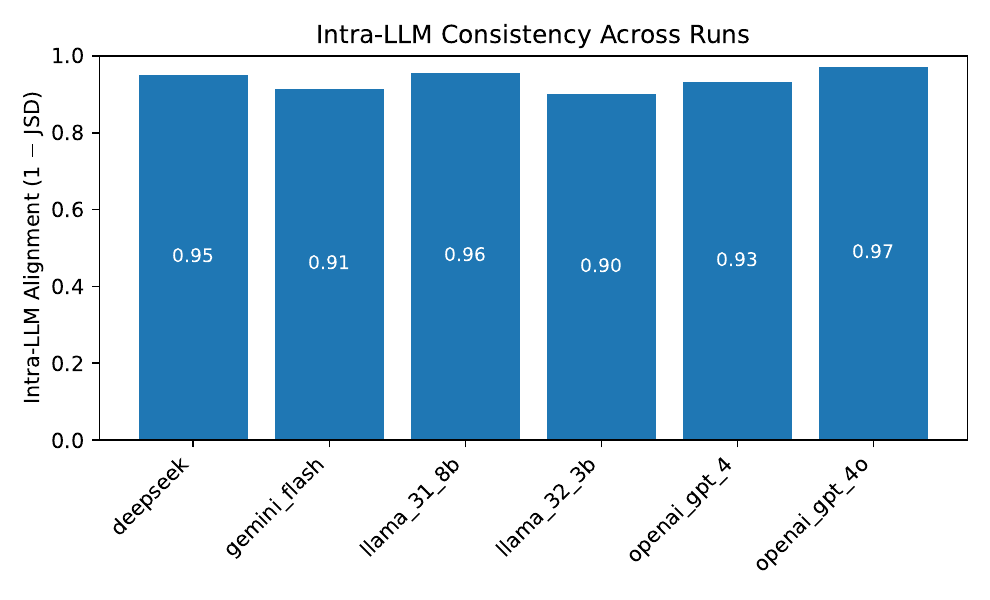}
    \caption{Intra-LLM stability across runs based on JSD, reported as $1-\mathrm{JSD}$ and averaged across all questions. Higher values indicate greater consistency of each model's generated response distributions across repeated runs.}
    \label{fig:exp2-intra}
\end{figure}

\paragraph{Intra-LLM variability}
\label{sec:exp2:intra}
\Cref{fig:exp2-intra} reports intra-LLM alignment across repeated
runs, measured as $1-\text{JSD}$ between distributions generated by the
same model. The results indicate that, for a fixed prompt and model,
independently generated runs converge to nearly identical aggregate
response distributions. Consequently, the divergence between LLM-generated and human survey distributions cannot be attributed to stochastic decoding variability, but reflects systematic differences in the distributions induced by the models.

\begin{figure}[t]
    \centering
    \includegraphics[width=1\linewidth]{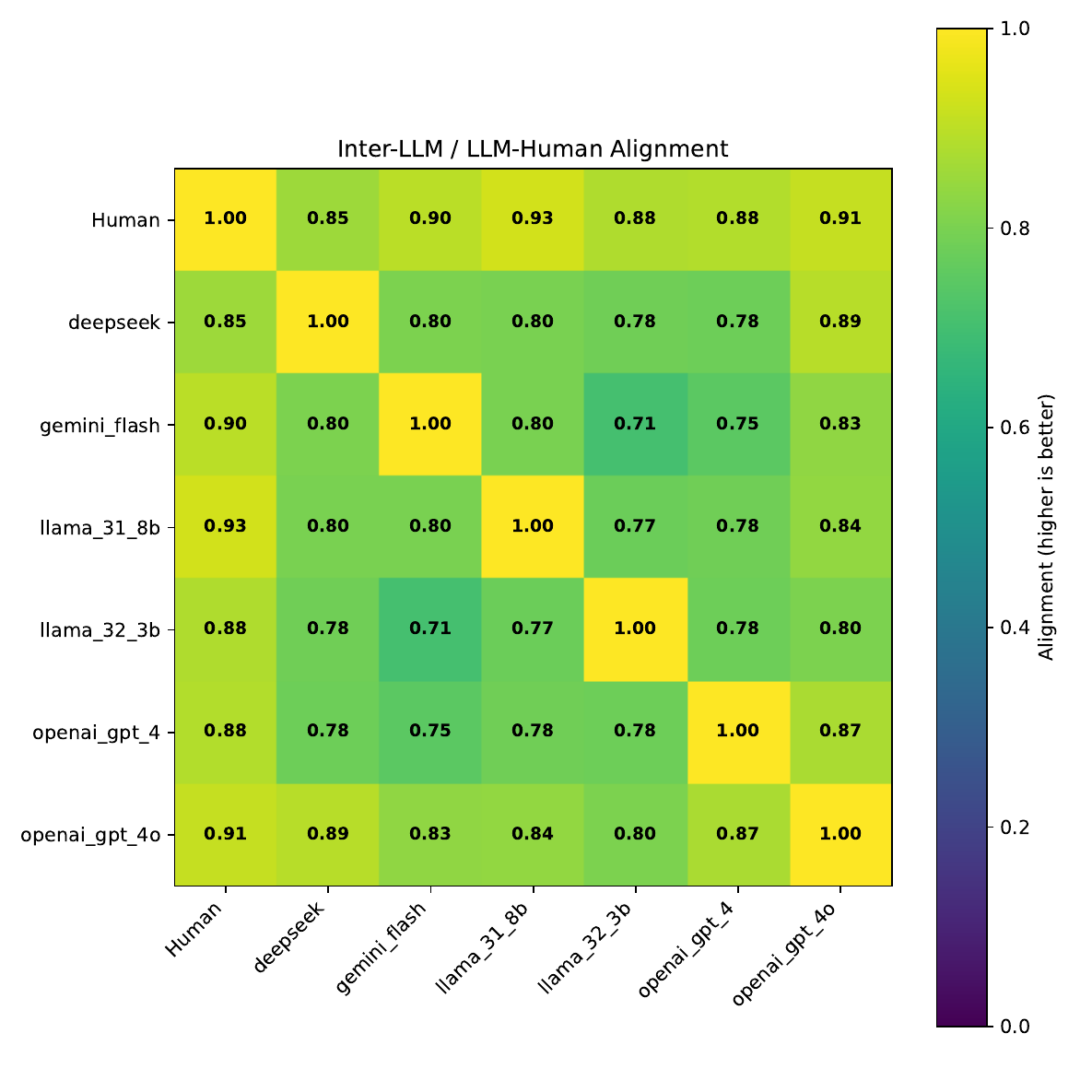}
    \caption{Pairwise distributional similarity between LLMs and the human reference based on Jensen--Shannon divergence, reported as $1-\mathrm{JSD}$ and averaged across all questions and runs. Higher values indicate greater similarity between response distributions. Numerical values inside the heatmap denote the aggregated mean similarity for each pair.}
    \label{fig:exp2-inter}
\end{figure}

\paragraph{Inter-LLM and LLM--human pairwise alignment}
To assess similarity across models, we compute pairwise JSD values for every model pair across all questions and runs. The resulting heatmap is shown in \Cref{fig:exp2-inter}, including the human reference distribution as an additional row and column.

Under this relative similarity measure, both inter-model alignment and LLM--human alignment appear high. This indicates that LLMs tend to produce broadly similar distributional shapes and that their outputs are close to the human reference in a pairwise sense. However, this result reflects relative similarity between distributions and does not imply accurate reproduction of the human response proportions, which we evaluate separately using absolute divergence measures.

\begin{figure}[t]
    \centering
    \includegraphics[width=0.9\linewidth]{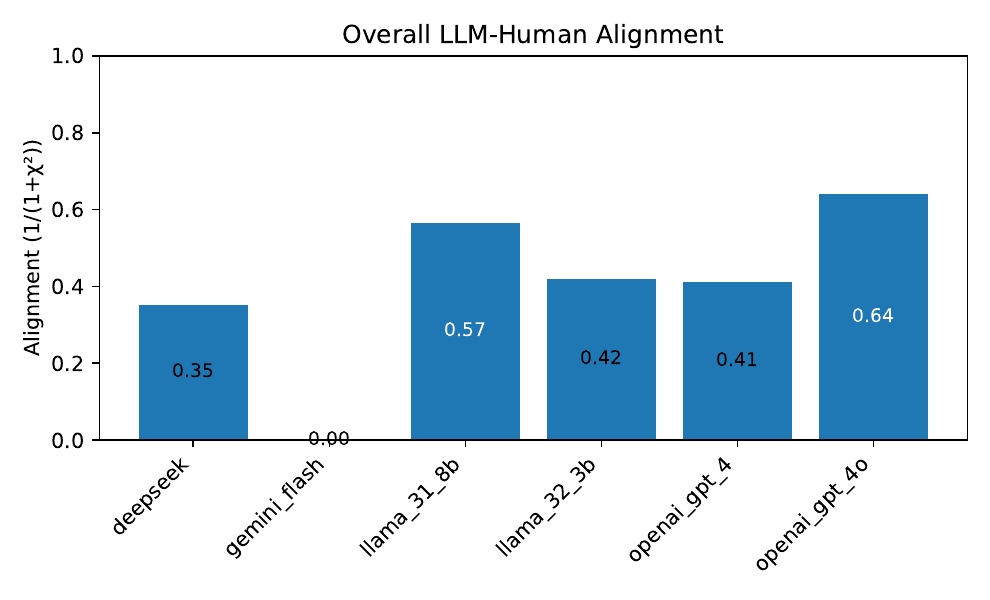}
    \caption{Overall LLM--human alignment based on $\chi^2$ divergence, reported as $1/(1+\chi^2)$ and aggregated across questions. Higher values indicate smaller absolute discrepancies between LLM-generated and human response distributions.}
    \label{fig:exp2-chi2}
\end{figure}

\paragraph{Overall LLM--human alignment (absolute fit)}
\Cref{fig:exp2-chi2} evaluates overall alignment between each
LLM-generated distribution and the human baseline using a chi-square
based score, reported as $1/(1+\chi^2)$. This metric produces a markedly different ranking. Thus, while pairwise similarity scores suggest closeness between models and the human reference, the chi-square evaluation reveals large absolute mismatches for most systems.

In summary, we report both Jensen--Shannon divergence (JSD) and $\chi^2$ divergence, for the LLM--human alignment, as they capture complementary aspects of agreement. JSD is bounded and emphasises overall distributional similarity, while $\chi^2$ strongly penalises relative errors in low-probability categories. Consequently, LLM--human alignment is consistently higher under JSD than $\chi^2$, reflecting that models reproduce the global shape of human responses while still over-allocating probability mass to rare options.

\paragraph{Overall findings}
Across all results, a consistent pattern emerges. LLMs are highly stable in the distributions they generate across runs and appear mutually similar under a pairwise alignment view. However, when evaluated using an absolute divergence measure that penalises distribution-level discrepancies, alignment with the human baseline degrades sharply for most models. These results show that internal consistency and pairwise similarity do not imply faithful reproduction of human survey distributions at the population level.

\subsection{Temporal survey simulation}
\label{sec:exp3}

We extend the distributional simulation of the previous setup to a multi-year setting. Using SANS SOC surveys from 2017--2019 and
2021--2025, we compare model-generated response distributions with the
published human percentages on a per-year basis. Alignment is measured using Jensen--Shannon divergence (JSD), reported as $1 - \mathrm{JSD}$ for interpretability, where higher values indicate closer alignment.

\begin{figure}[t]
    \centering
    \includegraphics[width=1\linewidth]{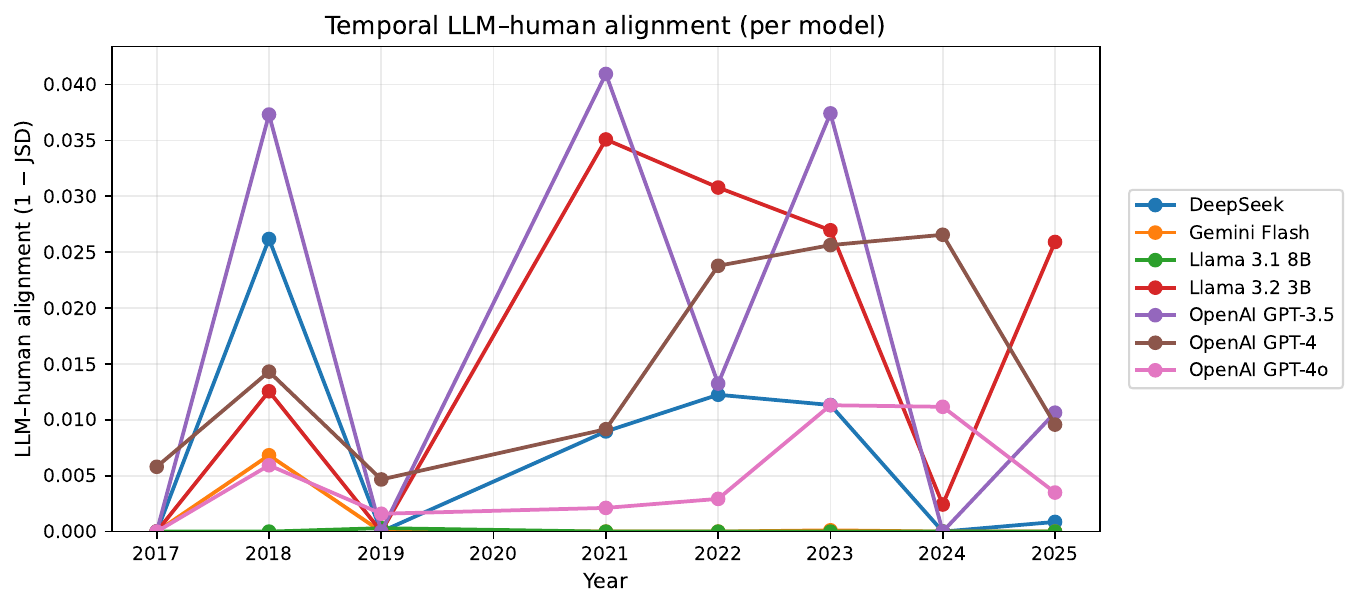}
    \caption{Temporal LLM--human alignment per model, reported as mean $1-\mathrm{JSD}$ over questions for each SANS SOC survey year.}
    \label{fig:temporal_alignment_per_model}
\end{figure}

\begin{figure}[t]
    \centering
    \includegraphics[width=1\linewidth]{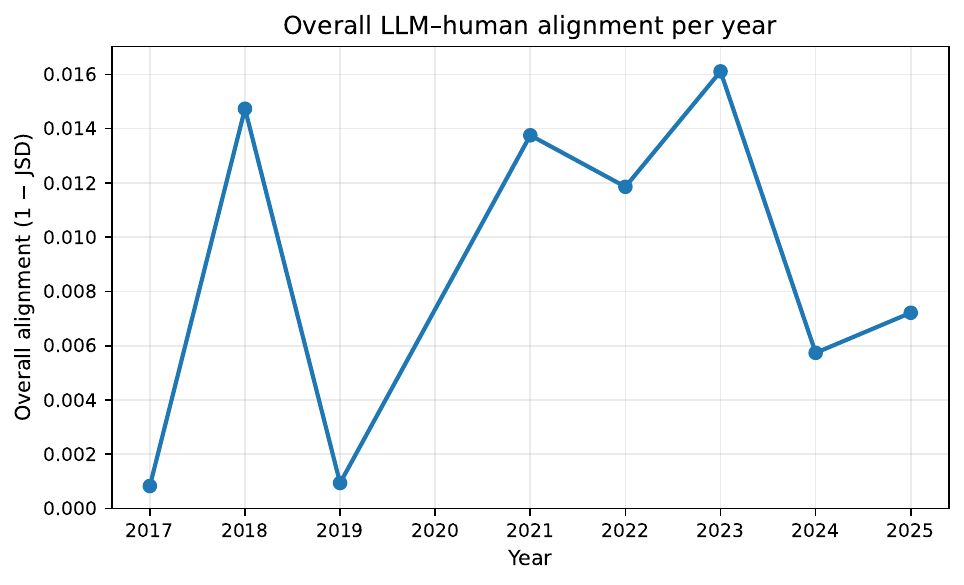}
    \caption{Ensemble LLM--human alignment per survey year, shown as mean $1-\mathrm{JSD}$ aggregated across all models.}
    \label{fig:temporal_alignment_ensemble}
\end{figure}

\paragraph{Per-model temporal alignment}
\Cref{fig:temporal_alignment_per_model} reports LLM--human alignment trajectories for each model across survey years, while \Cref{fig:temporal_alignment_ensemble} is the aggregated alignment scores across all models for each survey year. Across all evaluated models, alignment remains consistently low and tightly bounded over time. No model achieves close correspondence to the human reference distribution in any year. Importantly, alignment curves remain within a narrow band for all models, indicating that none meaningfully track temporal changes in practitioner response distributions.

\begin{figure}[t]
    \centering
    \includegraphics[width=1\linewidth]{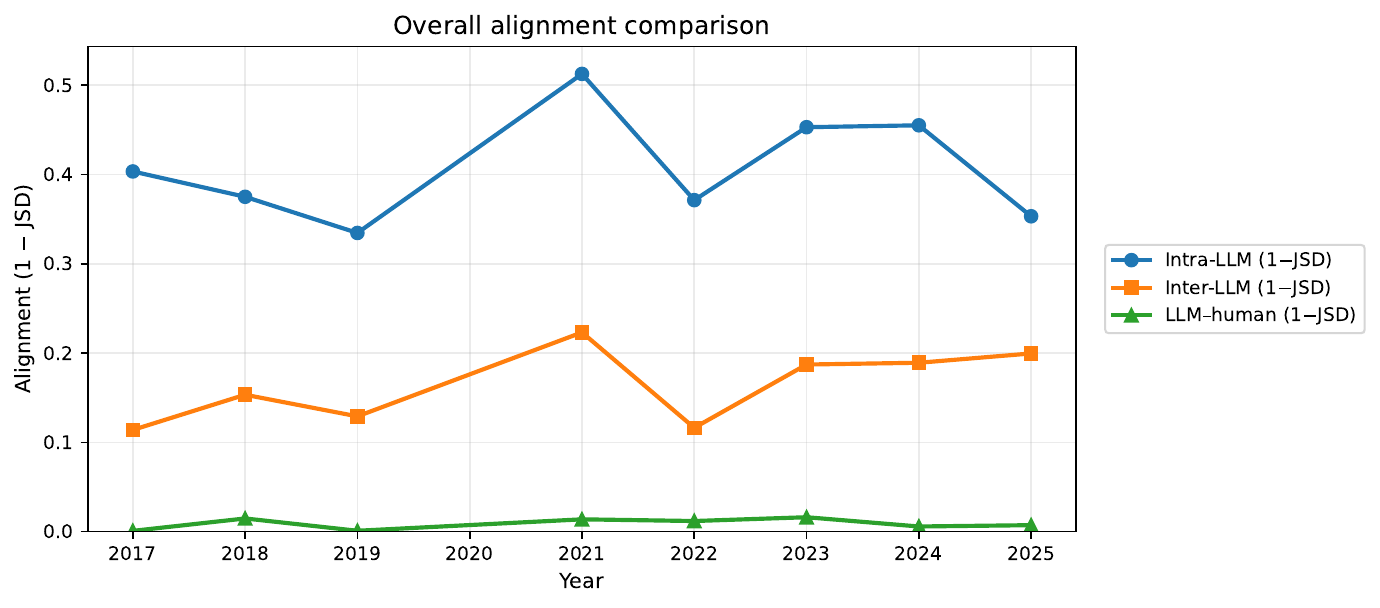}
    \caption{Comparison of intra-LLM, inter-LLM and LLM--human divergence over time, measured using mean $1-\mathrm{JSD}$ per survey year.}
    \label{fig:temporal_alignment_comparison}
\end{figure}

\paragraph{Divergence structure and cut-off-aware alignment}
\Cref{fig:temporal_alignment_comparison} compares intra-LLM divergence, inter-LLM divergence and LLM--human divergence across survey years. Across all years, intra-LLM divergence is consistently lowest, indicating that individual models reliably reproduce their own induced response distributions over repeated runs. Inter-LLM divergence is higher, reflecting systematic differences between model families, while LLM--human divergence is consistently highest. This ordering remains stable over time, confirming that temporal misalignment is dominated by disagreement between all models and the human reference rather than by stochastic sampling variability or cross-model noise.

\begin{figure}[t]
    \centering
    \includegraphics[width=1\linewidth]{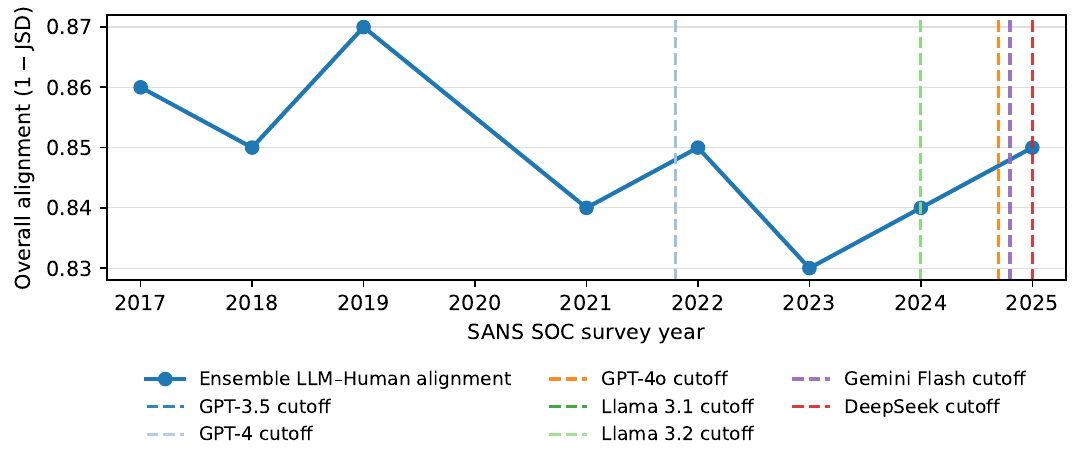}
    \caption{Overall (ensemble) LLM--human alignment over SANS SOC survey years, measured as mean $1-\mathrm{JSD}$. Vertical dashed lines indicate approximate training cutoff years.}
    \label{fig:cutoff_timeline_alignment}
\end{figure}

To assess whether this divergence pattern is related to model training horizons, we overlay approximate training cut-off years onto the ensemble LLM--human alignment trajectory in \Cref{fig:cutoff_timeline_alignment}. Alignment remains uniformly low both before and after cut-off points, with no observable improvement for pre-cut-off survey years nor systematic degradation for post-cutoff years. Taken together, these results indicate that while LLMs are internally stable and exhibit consistent inter-model differences over time, their divergence from human survey distributions is largely insensitive to training cut-offs and temporal distance from the nominal training window.

\begin{figure}[t]
    \centering
    \includegraphics[width=1\linewidth]{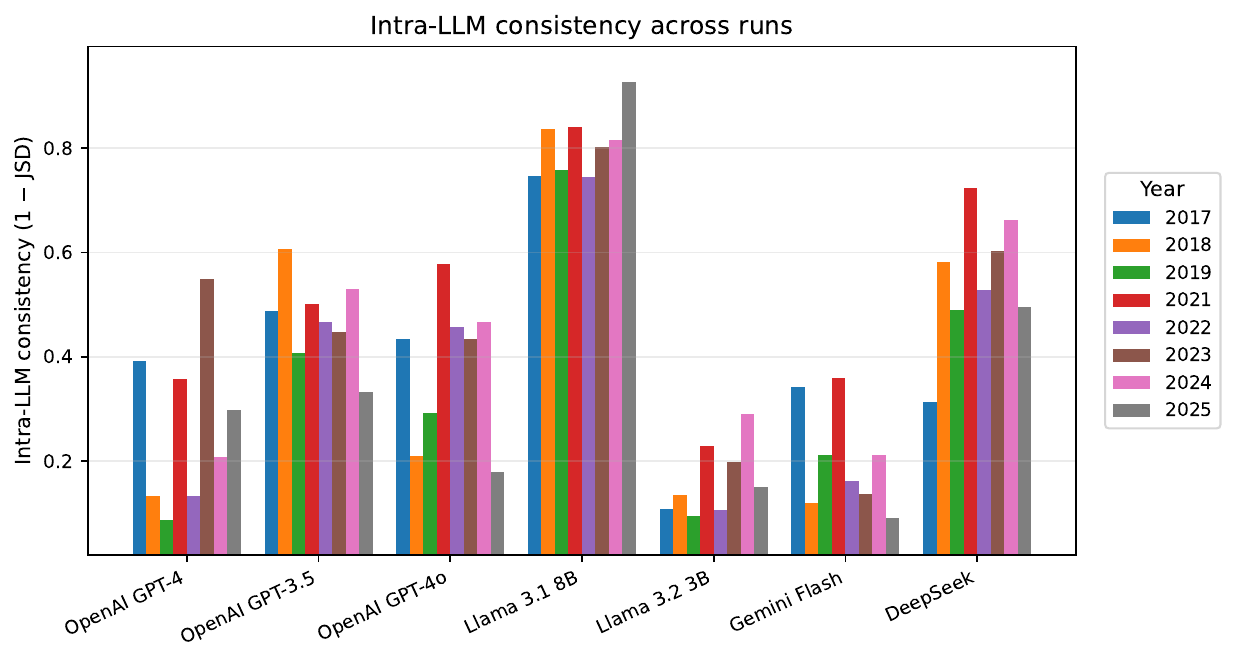}
    \caption{Intra-LLM consistency across survey years, measured as mean $1-\mathrm{JSD}$ between repeated runs of the same model. Higher values indicate greater stability of model-induced distributions over time.}
    \label{fig:intra_llm_temporal_consistency}
\end{figure}

\begin{figure}[t]
\centering
\includegraphics[width=1\linewidth]{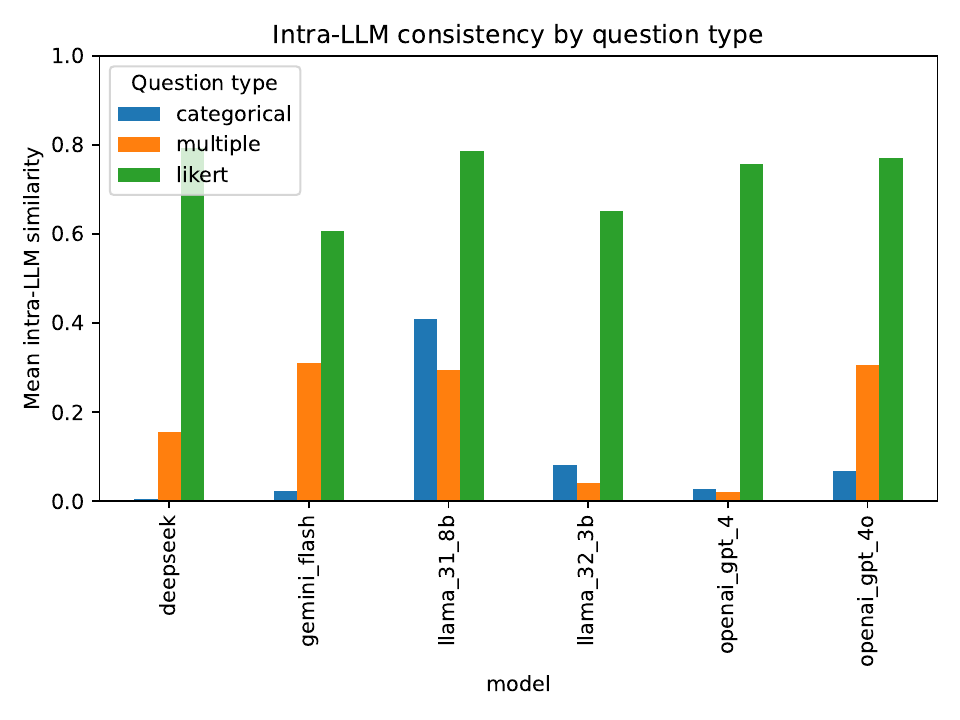}
\caption{Intra-LLM consistency across structured question types (categorical, multiple-choice, Likert).
Bars show the mean similarity across ten runs and six personas for each model; higher values indicate more stable repeated responses.}
\label{fig:intra-llm-merged}
\end{figure}

\paragraph{Temporal intra-LLM consistency}
Intra-LLM consistency remains high across all models and survey years (\Cref{fig:intra_llm_temporal_consistency}), confirming that each model reliably reproduces its induced response distributions under repeated sampling. This supports the conclusion that temporal misalignment with human survey data cannot be attributed to stochastic sampling variability, but instead reflects systematic differences between LLM-generated and human distributions. Detailed per-model consistency results are also provided in \Cref{fig:intra-llm-merged}.

\begin{figure}[t]
    \centering
    \includegraphics[width=1\linewidth]{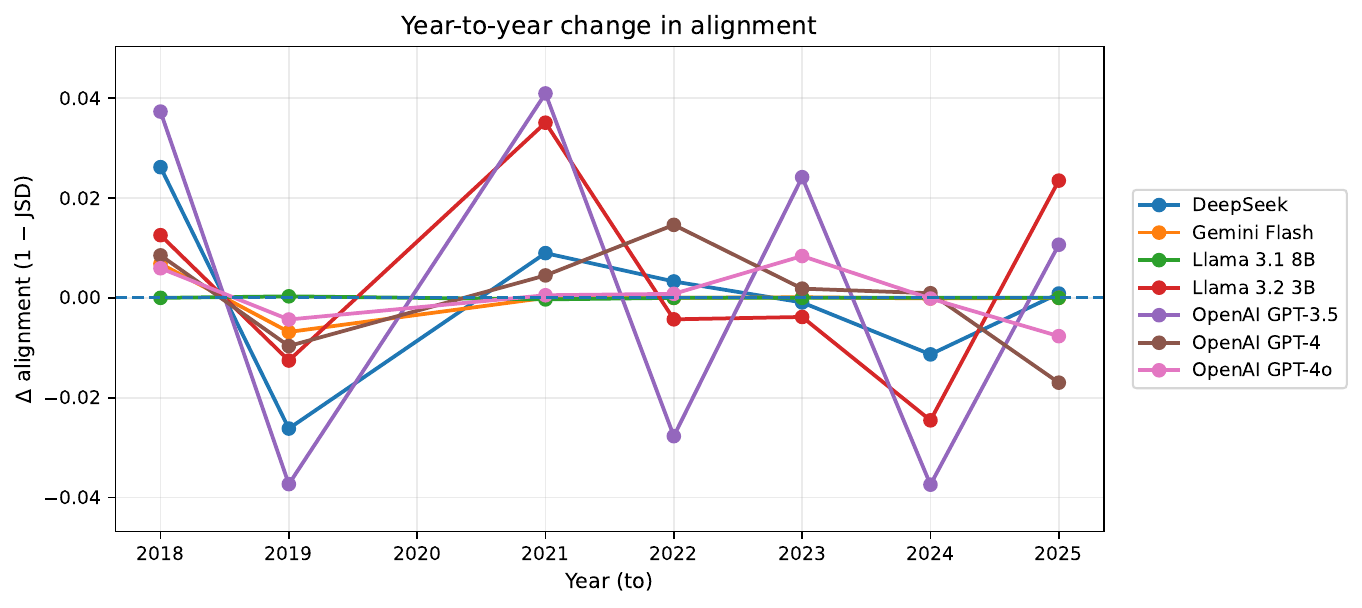}
    \caption{Year-to-year change in LLM--human alignment per model, measured as $\Delta(1-\mathrm{JSD})$ between consecutive years.}
    \label{fig:year_to_year_drift_models}
\end{figure}

\begin{figure}[t]
    \centering
    \includegraphics[width=1\linewidth]{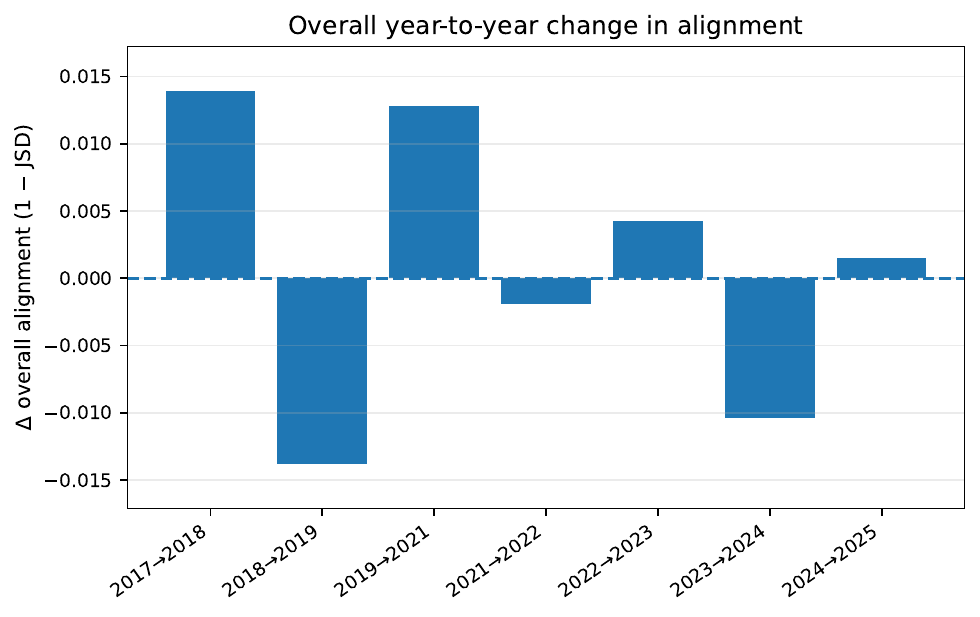}
    \caption{Year-to-year change in ensemble LLM--human alignment, computed as the change in mean $1-\mathrm{JSD}$ across all models.}
    \label{fig:year_to_year_drift_ensemble}
\end{figure}

\paragraph{Year-to-year drift}
We further examine temporal drift by computing year-to-year changes in alignment. \Cref{fig:year_to_year_drift_models,fig:year_to_year_drift_ensemble} show that changes in alignment are generally small and non-monotonic. Most transitions fall within a narrow range, with both positive and negative shifts observed across different models and years. There is no consistent trend towards improving or degrading alignment over time, suggesting irregular sensitivity to individual survey years rather than systematic temporal learning.

\paragraph{Overall findings}
Overall, the temporal analysis shows that LLMs generate stable but time-insensitive survey distributions. Alignment with humans is weak across all years, models and temporal perspectives. While models exhibit strong internal consistency and limited year-to-year fluctuation, they do not meaningfully track historical changes in practitioner responses, nor do they exhibit cut-off-dependent behaviour indicative of memorisation or temporal generalisation.

\subsection{Qualitative analysis}

We complement quantitative comparisons with qualitative observations of response patterns. Human experts frequently provided diverse and sometimes conflicting responses, reflecting differences in experience and organisational context. In contrast, LLM-generated responses tended to converge towards moderate and generalised answers, consistent with the observed central tendency bias.

Even when conditioned on distinct expert personas, models often produced responses aligned with generic best practices rather than reflecting clearly differentiated operational perspectives. This suggests that LLM simulations capture an averaged representation of expert knowledge rather than the variability inherent in real practitioners.

Additionally, LLM-generated distributions remained largely stable across survey years, failing to reflect temporal shifts observed in human data. This indicates that models reproduce static conceptual knowledge rather than evolving professional judgement. These qualitative differences help explain the observed quantitative misalignment and highlight limitations of LLMs as surrogate expert respondents.

\subsection{Answers to the research questions}
\label{sec:rq-summary}

We now explicitly summarise how our findings address the research questions posed in \Cref{methodology}.

\paragraph{RQ1: Individual-level alignment}
Our results show that LLMs generate internally coherent and moderately stable responses when simulating individual SOC experts. However, alignment with human expert responses remains weak. Intra-LLM consistency consistently exceeds LLM--human agreement, indicating that persona-conditioned simulations primarily reflect model-specific response patterns rather than expert-level judgements.

\paragraph{RQ2: Population-level distributions}
LLMs produce aggregate response distributions that are highly stable across repeated runs and often similar across models. Nevertheless, when evaluated using absolute divergence measures, these distributions diverge substantially from human survey data. This demonstrates that apparent distributional similarity does not imply faithful reproduction of practitioner response patterns.

\paragraph{RQ3: Stability and model dependence}
Across all experimental settings, LLM responses exhibit high intra-model stability but low inter-model agreement. However, this stability reflects internally consistent model behaviour rather than convergence towards human expert responses.

\paragraph{RQ4: Temporal generalisation}
LLM--human alignment remains consistently low across survey years and does not systematically vary with model training cut-offs. This indicates limited evidence of meaningful temporal generalisation and suggests that LLM-generated survey responses remain largely time-insensitive.

\paragraph{RQ5: Systematic differences and failure modes}
Taken together, these findings show that LLM-generated responses exhibit systematic differences from human expert data, including reduced variance, central tendency bias and homogenised response patterns.

\paragraph{RQ6: Methodological implications}
These findings indicate that LLMs are unsuitable as replacements for human expert respondents in cybersecurity surveys. However, their stability, controllability and scalability make them valuable as augmentation tools for survey piloting, instrument stress-testing and exploratory hypothesis generation when applied with appropriate methodological safeguards.
\FloatBarrier
\section{Limitations}
\label{limitations}

Our study has limitations that should be considered as opportunities for future research:  

\paragraph{Rapid evolution of LLMs}
As LLMs advance quickly, frameworks must adapt. Nevertheless, our framework is model-agnostic and evaluates methodological properties which remain relevant regardless of specific model versions.

\paragraph{Lack of comprehensive ground truth} Cybersecurity often lacks definitive ground truth on “expert” knowledge, complicating evaluation of AI–expert alignment. However, our goal is not to establish absolute correctness, but to compare LLM responses against real experts, which provides a valid and practical data pool for assessing surrogate suitability.

\paragraph{Zero-Shot prompting design} 
Our approach relies on zero-shot prompting to examine the models’ inductive biases in the absence of human example answers, while still providing task-relevant contextual information. Multi-shot or multi-turn prompting could elicit more stable or contextually grounded outputs but was out of this study's scope. This design allows us to evaluate the models’ inherent behaviour without prompt-specific optimisation, ensuring that our findings reflect their default capabilities rather than tuned performance.

\FloatBarrier
\section{Ethical considerations}
\label{ethics}

This study involves human expert participants and synthetic responses generated by LLMs, raising ethical considerations related to human subjects, representation and potential misuse. We evaluated these considerations using a stakeholder-based ethics analysis following the principles of the Menlo Report \cite{bailey2012menlo}.

Human expert participation was conducted with appropriate institutional review board (IRB) approval. Participation was voluntary, no monetary compensation was provided and no personally identifiable information was collected or stored. Reported demographic information is presented only in aggregated form to prevent re-identification. The study design minimised risks to participants by focusing on professional practices rather than sensitive personal information.

A central ethical consideration in this work is the potential misinterpretation or misuse of LLM-generated responses. Because synthetic outputs may appear plausible or authoritative, there is a risk that such responses could be mistaken for genuine expert opinion or used to replace human expertise. Our findings explicitly demonstrate systematic differences between LLM-generated and human expert responses, and we emphasise that LLMs should not be used as substitutes for expert elicitation in security research. Instead, we position LLM-generated responses strictly as methodological tools for piloting, hypothesis generation or survey design support.

Our work also raises broader concerns regarding representation and bias. LLM-generated responses reflect patterns learned from training data and may obscure minority viewpoints, reduce variability or reinforce dominant perspectives. These limitations could lead to misleading conclusions if synthetic responses are treated as representative of real experts. We therefore advocate for careful validation and transparent reporting when using LLMs in human-centred research.

Finally, although this study did not involve sensitive operational data, the use of LLMs in security contexts may introduce privacy and confidentiality risks if applied improperly. Researchers should avoid submitting sensitive information to external models without appropriate safeguards.

Overall, we frame LLM-generated survey responses as supplementary methodological tools rather than replacements for human expertise, and we highlight the importance of maintaining human-centred validation in cybersecurity research.

\subsection{Stakeholders}
The primary stakeholders in this work include:

\begin{itemize}
    \item \textbf{Human experts}, whose professional expertise could be compared against or supplemented by LLM outputs.
    \item \textbf{Researchers and practitioners} in cybersecurity, who may adopt, interpret or extend our methods.
    \item \textbf{Organisations and decision-makers}, who could use results to inform security operations or policies.
    \item \textbf{The broader public and society}, who are indirectly affected by the trustworthiness and equity of cybersecurity practices shaped by such research.
\end{itemize}

\subsection{Potential harms and mitigations}

\paragraph{Risk of replacing human expertise.}
\textbf{\emph{Harm}:} LLMs may produce expert-like responses that unintentionally devalue or replace genuine human expertise.
\textbf{\emph{Mitigation}:} We frame LLMs strictly as supplementary tools to extend limited human input, emphasising that human expertise remains indispensable for decision-making.

\paragraph{Accountability and trust.}
\textbf{\emph{Harm}:} The opaque reasoning of LLMs complicates accountability when their outputs are used.
\textbf{\emph{Mitigation}:} We disclose when and how AI-generated responses are included, interpret them only within methodological limits and avoid presenting them as authoritative judgements.

\paragraph{Responsible use of AI-generated insights.}
\textbf{\emph{Harm}:} Applying LLM outputs directly in operational contexts (e.g. incident response) could lead to harmful outcomes.
\textbf{\emph{Mitigation}:} We explicitly position our work as exploratory and hypothesis-generating, requiring human-in-the-loop validation for any practical use.

\paragraph{Data privacy and confidentiality.}
\textbf{\emph{Harm}:} Although our study did not involve sensitive or personal data, deploying LLMs with sensitive prompts in real-world settings could risk unintended disclosure of confidential information, particularly when interacting with third-party services.
\textbf{\emph{Mitigation}:} We stress the need for explicit safeguards against such disclosures in applied contexts, including careful handling of data, transparency in model use and strict avoidance of sensitive inputs.

\paragraph{Misinterpretation and overgeneralisation.}
\textbf{\emph{Harm}:} Stakeholders might assume LLM-generated insights represent consensus expert views or generalise across populations.
\textbf{\emph{Mitigation}:} We highlight methodological limitations and communicate clearly that synthesised responses cannot replace diverse expert perspectives. We emphasise that LLM outputs must not be mistaken for consensus or population-wide views, and we frame them strictly as exploratory tools to avoid overclaims or bias.

\paragraph{Dual use and equity concerns.}
\textbf{\emph{Harm}:} Techniques for simulating expertise could be misused (e.g. to generate attacker strategies or deceptive content). In addition, LLMs are predominantly trained on English and Western-centric sources, which risks amplifying certain perspectives while marginalising others, thereby excluding non-Western viewpoints from consideration.
\textbf{\emph{Mitigation}:} We acknowledge these risks explicitly and advocate for careful, equitable and transparent application of such methods. Awareness of these limitations is necessary to promote responsible use and to avoid reinforcing existing imbalances in expertise representation.

\subsection{Ethical principles and decision}

\begin{itemize}
    \item \textbf{Beneficence}: the potential benefits of exploring how LLMs can supplement limited expert survey data outweigh the manageable risks, given our mitigations.
    \item \textbf{Respect for persons}: we respect the autonomy and dignity of human experts by framing LLMs as augmentative rather than substitutive.
    \item \textbf{Justice}: we consider distributional impacts, noting risks of excluding under-represented perspectives, and we document these risks openly.
    \item \textbf{Respect for law and public interest}: no legal or contractual obligations were violated; institutional ethics review approval was obtained.
\end{itemize}

Given these considerations, we concluded that proceeding with and publishing this research is ethically justified. Our study provides transparency about both the capabilities and the limitations of LLMs in cybersecurity contexts, enabling informed discourse while mitigating foreseeable harms. A complementary discussion of methodological constraints is provided in \Cref{limitations}.
\FloatBarrier
\section{Conclusion}
\label{Conclusion}

Cybersecurity research depends fundamentally on expert knowledge to understand practitioner workflows, decision-making and operational realities. In this work, we presented a methodological framework for evaluating large language models as surrogate participants in expert security surveys. By applying this framework across individual, aggregate and temporal settings, we showed that while LLM-generated responses are internally stable and reproducible, they do not reliably capture the variability or distributional characteristics of real expert judgements. These differences manifest as reduced variance, central tendency bias and a failure to capture the diversity of expert perspectives.

These findings have important implications for human-centred security research. Our results demonstrate that synthetic responses reflect model-specific patterns rather than authentic practitioner perspectives, limiting their validity as substitutes for human participants. However, their controllability and scalability make them useful methodological tools for piloting survey instruments, exploring hypothetical scenarios and stress-testing research designs.

More broadly, this work provides methodological guidance for evaluating LLM-generated survey responses and highlights the importance of preserving human expertise in cybersecurity research. We hope this framework supports researchers in responsibly integrating LLMs into survey-based studies while maintaining rigorous standards for human-centred validity.
\FloatBarrier

\section*{Conflicts of interest}
The authors declare that they have no competing interests.

\section*{Funding}
No specific financial support was received for this work.

\section*{Data availability}
The data underlying this article are available at \url{https://github.com/desgiar/llm-survey-artifacts}.

\section*{Author contributions}
D.G. conceptualisation, formal analysis, software, data curation, investigation, methodology, validation, visualisation, writing---original draft, writing---review \& editing.
R.M. supervision, methodology, writing---review \& editing.
T.F.B. conceptualisation.
V.L. conceptualisation, supervision, methodology, writing---review \& editing.
J.K. supervision, writing---review \& editing.

\section*{Acknowledgments}
The authors thank the anonymous reviewers for their valuable suggestions.
We acknowledge the use of GPT-5.5 for drawing \Cref{fig:flowchart_soups} and Claude Opus 4.7 for polishing the paper.


\appendix

\section{Survey instrument}
\label{app:survey}

This appendix presents the full survey instrument used in our expert study on AI adoption in SOCs. The survey was administered to cybersecurity professionals working in SOC environments. The instrument consisted of five sections: demographics, SOC characteristics, processes and tooling, AI adoption, and professional perceptions.

Unless otherwise stated, closed questions used multiple-choice or Likert-scale formats.

\subsection{Demographics}

\begin{itemize}

\item \textbf{Job title}
\begin{itemize}
\item Security Analyst (Tier 1)
\item Security Investigator (Tier 2)
\item Security Engineer
\item Security Architect
\item Forensic Analyst
\item Threat Hunter
\item Penetration Tester
\item Data Scientist
\item AI Expert
\item SOC Manager
\item Product/Project Manager
\end{itemize}

\item \textbf{Years of experience in this role}
\begin{itemize}
\item < 1 year
\item 1--3 years
\item 3--5 years
\item 5--10 years
\item 10+ years
\item Prefer not to say
\end{itemize}

\item \textbf{Self-evaluation of expertise (1=Beginner, 5=Expert)}

\item \textbf{Level of Education}
\begin{itemize}
\item High School
\item Technical/Vocational training
\item Certification/Diploma
\item Bachelor's Degree
\item Master's Degree
\item Doctoral Degree
\end{itemize}

\end{itemize}

\subsection{SOC characteristics}

\begin{itemize}

\item \textbf{Organisation size}
\begin{itemize}
\item <10
\item 11--49
\item 50--99
\item 100--499
\item 500--999
\item 1,000--5,000
\item 5,000--10,000
\item >10,000
\item Unknown
\end{itemize}

\item \textbf{SOC team size (numeric)}

\item \textbf{SOC type}
\begin{itemize}
\item Industrial
\item Academic
\item Governmental
\item Prefer not to say
\end{itemize}

\item \textbf{Country of SOC}

\item \textbf{SOC model}
\begin{itemize}
\item Dedicated SOC
\item Managed SOC
\item Hybrid SOC
\item Command SOC
\item SOC/NOC
\item Uncertain
\end{itemize}

\item \textbf{SOC maturity level}
\begin{itemize}
\item Non-existent
\item Initial
\item Managed
\item Defined
\item Quantitatively managed
\item Optimising
\end{itemize}

\item \textbf{Does SOC operate 24/7?}

\end{itemize}

\subsection{SOC processes and workload}

\begin{itemize}

\item \textbf{Number of alerts received daily}

\item \textbf{Percentage of alerts}
\begin{itemize}
\item Investigated
\item Remediated
\item Escalated
\item False positives
\end{itemize}

\item \textbf{Capabilities present in SOC}
\begin{itemize}
\item Detection
\item Investigation
\item Response
\item Prevention
\item Remediation
\item Prediction
\item Reporting
\item Automation
\end{itemize}

\item \textbf{Level of automation in SOC functions}
\begin{itemize}
\item Fully automated (AI)
\item Fully automated (rule-based)
\item Partially automated
\item Not automated
\end{itemize}

\end{itemize}

\subsection{AI Usage and Tooling}

\begin{itemize}

\item \textbf{Have you used AI methods in SOC?}

\item \textbf{Types of AI used}
\begin{itemize}
\item Supervised learning
\item Unsupervised learning
\item Neural networks
\item Deep learning
\item UEBA
\end{itemize}

\item \textbf{Likert-scale statements (1--5)}

Examples:

\begin{itemize}
\item AI tools improve threat detection
\item AI tools reduce false positives
\item AI tools affect workload
\item AI tools are difficult to integrate
\item Explainability is important
\item AI tools can improve SOC performance
\end{itemize}

\end{itemize}

\subsection{Challenges and perceptions}

\begin{itemize}

\item Rate how challenging the following are (1--5):

\begin{itemize}
\item Incident response
\item Threat detection
\item Threat hunting
\item Data analysis
\end{itemize}

\end{itemize}

\subsection{Professional perceptions}

Likert-scale questions (1--5):

\begin{itemize}
\item I trust AI tools
\item AI tools improve productivity
\item AI tools increase security effectiveness
\item AI tools present risks
\item Human analysts outperform AI
\end{itemize}

\subsection{Tool usage}

\begin{itemize}
\item Tools used:
\begin{itemize}
\item SIEM
\item EDR
\item XDR
\item SOAR
\item IDS/IPS
\item Threat intelligence platforms
\end{itemize}
\end{itemize}

\end{document}